\documentclass[twocolumn,astrosymb,tighten,twocolappendix,trackchanges]{aastex631}
\usepackage{longfigure}

\shorttitle{}
\shortauthors{}
\turnoffedit

\graphicspath{{./}{figures/}}

\begin{document}

\title{Searching for Radio Transients with Inverted Spectra in Epoch 1 of VLASS and VCSS, and Identification of a Sample of Candidate Relativistic Nuclear Transients}

\correspondingauthor{Yuyang Chen}
\email{yuyangf.chen@mail.utoronto.ca}

\author[0000-0002-8804-3501]{Yuyang Chen}
\affiliation{David A. Dunlap Department of Astronomy \& Astrophysics, University of Toronto, Toronto, ON M5S 3H4, Canada}
\affiliation{Dunlap Institute for Astronomy \& Astrophysics, University of Toronto, Toronto, ON M5S 3H4, Canada}

\author[0000-0002-3382-9558]{B. M. Gaensler}
\affiliation{Dunlap Institute for Astronomy \& Astrophysics, University of Toronto, Toronto, ON M5S 3H4, Canada}
\affiliation{David A. Dunlap Department of Astronomy \& Astrophysics, University of Toronto, Toronto, ON M5S 3H4, Canada}
\affiliation{Department of Astronomy and Astrophysics, University of California Santa Cruz, Santa Cruz, CA 95064, USA}

\author[0000-0001-6812-7938]{Tracy Clarke}
\affiliation{U.S. Naval Research Laboratory, 4555 Overlook Ave. SW, Washington, DC 20375, USA}

\author[0000-0002-5187-7107]{Wendy Peters}
\affiliation{U.S. Naval Research Laboratory, 4555 Overlook Ave. SW, Washington, DC 20375, USA}

\author[0000-0003-3272-9237]{Emil Polisensky}
\affiliation{U.S. Naval Research Laboratory, 4555 Overlook Ave. SW, Washington, DC 20375, USA}

\author[0000-0002-7329-3209]{Kovi Rose}
\affiliation{Sydney Institute for Astronomy, School of Physics, University of Sydney, NSW, 2006, Australia}
\affiliation{CSIRO Space and Astronomy, PO Box 76, Epping, NSW, 1710, Australia}

\begin{abstract}

For radio transients, an inverted spectrum (defined as $\alpha > 0$ for a power law spectrum $S_{\nu}\propto \nu^{\alpha}$) constrains physical properties, which in principle can be a useful criterion for selecting specific targets of interest in a transient search. To test and develop this concept, we have searched epoch 1 of the Very Large Array Sky Survey (VLASS; 3.0\,GHz) and the VLITE Commensal Sky Survey (VCSS; 340\,MHz) for radio transients with inverted spectra. We discover a sample of 21 inverted-spectra transient candidates that are not associated with cataloged active galactic nuclei (AGNs). To the best of our knowledge, three of our candidates have recently been reported by others as radio transients, but none have reported transient counterparts at other wavelengths. We find that our candidates evolve slowly over years and show either highly inverted spectra or peaked spectra over $\sim$1--3\,GHz. Within our sample, nine candidates are matched to optical centers of galaxies and have estimated radio spectral luminosities of $L_{3.0\mathrm{GHz}}\sim10^{30}-10^{33}\,\mathrm{erg}\,\mathrm{s}^{-1}\,\mathrm{Hz}^{-1}$. Based on the observed properties, we find the most plausible transient classification for our candidates to be relativistic tidal disruption events. However, it is difficult to rule out variable or transient AGNs with highly inverted spectra. Upon examining physical constraints, we confirm that mainly relativistic transients (on-axis or off-axis) with equipartition energy $E_{\mathrm{eq}}\gtrsim10^{49}-10^{53}\,\mathrm{erg}$ are expected from searching VLASS and VCSS based on inverted spectra. The obtainable physical constraints, however, can be weak due to degeneracy introduced by viewing angle.

\end{abstract}

\section{Introduction} \label{sec:intro}

Transient phenomena such as supernovae (SNe), tidal disruption events (TDEs), and gamma-ray burst (GRB) afterglows are known to produce slow-evolving radio synchrotron emission from shock-accelerated particles \citep[e.g.,][]{Chandra2012,Alexander2020,Bietenholz2021RadioSN}. These phenomena therefore also manifest as slow\footnote{We use the term ``slow'' to differentiate transients that evolve over days to years from unique populations of fast radio transients that have durations of milliseconds to seconds, such as fast radio bursts \citep[e.g., see][]{Pietka2015}} transients at radio frequencies. Such radio synchrotron emission contains valuable information regarding the energetics, surrounding environment, and emission size of the transient, offering a unique perspective of the physical processes involved \citep[][]{Chevalier1998,Granot&Sari2002,BarniolDuran2013}. 

In the past, the study of (slow) transients at radio frequencies relied heavily on follow-up observations of events discovered by optical and X-ray telescopes. Now, new generations of wide-field radio sky surveys are enabling systematic discoveries of radio transients. To provide some examples, the Australian SKA Pathfinder \citep[ASKAP;][]{Johnston2007ASKAP,Hotan2021ASKAP} has been conducting surveys such as the Rapid ASKAP Continuum Survey \citep[RACS;][]{McConnell2020RACSI,Hale2021RACSIIcatalog,Duchesne2023RACSmidCatalog} and the Variable and Slow Transients survey \citep[VAST;][]{Murphy2013VAST,Murphy2021VAST} over selected frequencies spanning $\nu\simeq0.7$--$1.8\,\mathrm{GHz}$. In RACS and the VAST pilot survey, discoveries include a number of GRB afterglow candidates \citep[][]{Leung2021,Leung2023}, highly variable sources near the Galactic Center including candidate Galactic Center Radio Transients \citep[][]{Wang2022}, radio counterparts to known classical novae \citep[][]{Gulati2023}, numerous minute-timescale variable and transient radio sources \citep[][]{Wang2023}, flaring radio stars \citep[][]{Pritchard2024}, a sample of radio-classified TDEs \citep[][]{Dykaar2024}, radio counterparts to optically-discovered TDEs \citep[][]{Anumarlapudi2024}, and a number of SNe with re-brightening radio counterparts at late times \citep[][]{Rose2024}. 

The MeerKAT telescope \citep[][]{JonasMeerKAT2016} has been conducting surveys dedicated to variable and transient searches such as ThunderKAT \citep[][]{Fender2016ThunderKAT}. Searches in MeerKAT data and in fields of ThunderKAT have led to discoveries such as stellar transients \citep[][]{Driessen2020MKTJ1704,Andersson2022MKTJ1746,Driessen2022EXO0408}, scintillating active galactic nuclei \citep[AGNs;][]{Driessen2022GX339-4Field}, as well as upper limits at various timescales and sensitivities in specific fields \citep[][]{Rowlinson2022MAXIJ1820Field,Chastain2023GRBField,Chastain2024GRBSNField}. In addition, a number of studies have developed and explored new transient search methods using MeerKAT observations \citep[][]{Andersson2023MeerKATCitizenScience,Andersson2024VolunteerClassification,Fijma2024}.

The Low Frequency Array \citep[LOFAR;][]{vanHaarlem2013LOFAR} has been conducting surveys such as the LOFAR Two-metre Sky Survey \citep[LoTSS][]{Shimwell2017LoTSS,Shimwell2019LoTSSDR1,Shimwell2022LoTSSDR2} at 144\,MHz and the LOFAR Low Band Antenna Sky Survey \citep[LoLSS][]{deGasperin2021LoLSS,deGasperin2023LoLSSDR1} at 54\,MHz. Searches of LoTSS have yielded an upper limit on the density of transients with a timescale of $\sim$2--9\,yr \citep[][]{deRuiter2021} and one minute-timescale transient source \citep[][]{deRuiter2023}.

The Karl G. Jansky Very Large Array \citep[VLA;][]{Perley2011JVLA} has previously conducted the Caltech-NRAO Stripe 82 Survey \citep[CNSS;][]{Mooley2016CNSS} at 3.0\,GHz over the $\sim$270\,deg$^2$ Stripe 82 region during 2012--2014. The CNSS was a survey dedicated to transient search and led to discoveries such as stellar flares \citep[][]{Mooley2016CNSS}, the first radio-discovered TDE \citep[][]{Anderson2020}, and AGNs that transitioned from radio-quiet to radio-loud \citep[][]{Mooley2016CNSS,KunertBajraszewska2020,Wolowska2021}. 

Since 2017, the VLA has been conducting the VLA Sky Survey \citep[VLASS;][]{Lacy2020VLASS} at 3.0\,GHz over the entire sky North of $-40^{\circ}$ declination. Currently, VLASS is scheduled to complete in 2026 with four epochs of observations (the fourth being a half-epoch) separated by $\sim$1000\,d, and is therefore ideal for detecting slow radio transients that evolve over years to decades. Transient searches in VLASS are actively ongoing, and studies so far have showcased promising results. In VLASS, individually reported discoveries include a decades-long off-axis GRB afterglow \citep[][]{Law2018,Marcote2019,Mooley2022}, the first known merger-triggered SN \citep[][]{Dong2021}, an emerging pulsar wind nebula harboring a highly magnetized decades-old neutron star \citep[][]{Dong2023}, and a number of TDE candidates \citep[][]{Ravi2022,Somalwar2022} including possibly the first radio-selected relativistic TDE \citep[][]{Somalwar2023J0243}. Systematic searches of VLASS have also revealed populations of AGNs and galaxies that transitioned from radio-quiet to radio-loud \citep[][]{Nyland2020,Zhang2022,Kunert-Bajraszewska2025}, long-lasting radio counterparts of a number of SNe \citep[][]{Stroh2021}, a sample of radio-selected TDEs that tends to have properties differing from optically-selected TDEs \citep[][]{Somalwar2023TDEI,Somalwar2023TDEII,Somalwar2023pTDE}, and a sample of solar-type stars that display transient or variable radio emission \citep[][]{Davis2024}. Additionally, a commensal instrument on the VLA -- the VLA Low-band Ionosphere and Transient Experiment \citep[VLITE;][]{Clarke2016VLITE,Polisensky2016,Polisensky2024} -- has been conducting the VLITE Commensal Sky Survey \citep[VCSS;][]{Peters2021VCSS} simultaneously with VLASS at 340\,MHz. In VCSS, three radio flares have been detected from hot magnetic stars \citep[][]{Polisensky2023}. 

Evidently, these surveys provide an unprecedented view of the dynamic radio sky. However, radio surveys themselves provide limited information. In addition to the discovery, extensive follow-up observations are most often required to characterize the properties of a radio-identified transient. For example, targeted multi-band radio observations are usually necessary for extracting the radio spectral energy distribution (SED) and for understanding the emission mechanism \citep[such as the characteristic broken power law profile representing a synchrotron spectrum;][]{Sari1998,Granot&Sari2002}, and optical imaging and spectroscopy are often crucial for classification and host association (also important for distance determination in many cases). 

Since follow-up observations require a range of additional resources, they cannot be performed for every transient and are typically dedicated to the most unique and scientifically interesting discoveries. As a result, the selection of transients worthy of follow-up is an important step for a systematic search. Such selection usually revolves around specific properties. Some simple and effective approaches adopted in the literature involve selecting transients based on brightness \citep[e.g., the extraordinarily bright relativistic TDE candidate discovered by][]{Somalwar2023J0243}, timescale \citep[e.g., the decades-long off-axis GRB afterglow discovered by][]{Law2018}, or specific classifications of interest \citep[e.g., AGNs that appear as radio transients, as discovered by][]{Nyland2020}.

In principle, the spectral index $\alpha$ is another property that can be useful for this selection process, which characterizes the steepness of a power law spectrum (we adopt the definition $S_\nu \propto \nu^{\alpha}$ for a power law throughout this study). Certain types of radio sources are known to have distinctive distributions of spectral index. For example, pulsars are known to have steep negative spectral indices \citep[$\alpha\sim-1.6$;][]{Jankowski2018} while certain blazars and X-ray binaries are known to produce relatively flat radio spectra \citep[$\alpha\sim0$;][]{Padovani2016,HjellmingJohnston1988}. Therefore, spectral index information can be useful for classification schemes and the selection of specific types of objects \citep[e.g.,][]{Riggi2024}. 

Certain properties may also be inferred from spectral indices. In particular, a positive spectral index ($\alpha>0$) can be an indication of absorption \citep[][]{RybickiLightman1979}, including synchrotron self-absorption (SSA) and free-free absorption (FFA). In the radio, a spectrum with a positive spectral index is sometimes referred to as an inverted spectrum because the typical spectral index of radio sources is negative \citep[e.g., see spectral index distributions derived from VLASS;][]{Gordon2021VLASS}. An inverted spectrum places lower limits on the peak/turnover frequency and flux density, and constrains physical properties such as the emission size \citep[][]{KellermannPaulinyToth1981}, which, for an expanding outflow (assuming negligible external absorption), restricts the dynamical age (i.e., a high-frequency peak implies a small size and a young dynamical age). This concept has been used to argue for the youth of some compact radio-loud AGNs that have inverted spectra below some $\sim$GHz spectral peak \citep[under the assumption of SSA;][]{Shklovsky1965,Snellen1998,An&Baan2012}. The same concept applies to transients that in many cases produce radio emission through an expanding shock. In particular, for radio synchrotron transients, the peak frequency is expected to continuously decline and the spectral index (at $\sim$GHz frequency) should transition from a steep positive value (such as $\alpha=2.5$ for SSA) to a steep negative value ($\alpha\sim -1.0$) as the transient ages and becomes optically thin \citep[][]{Sari1998,Granot&Sari2002}. Consequently, an inverted spectrum for a transient could also imply some degree of youth. Therefore, a positive spectral index (or an inverted spectrum) can be useful for constraining physical properties during the selection of transients and is potentially a viable criterion for selecting dynamically young transients.

In practice, spectral information has not always been accessible from radio surveys. Usually, spectral indices are derived from comparing different surveys at different frequencies taken at different times \citep[e.g.,][]{deGasperin2018indexmap}, which are inaccurate for transient sources due to spectral evolution. In many of the new generation radio surveys, in-band spectral information will be provided as part of the data products \citep[][]{Shimwell2017LoTSS,Lacy2020VLASS,McConnell2020RACSI}, which will undoubtedly be beneficial for studying transients. However, in-band information has some limitations such as limited spectral coverage, large uncertainty, and the need for additional data processing resulting in delay in its release. Currently, useful spectral information is still largely unavailable in released catalogs from ongoing radio surveys. At the moment, none of the reported systematic transient searches in ongoing radio surveys make use of spectral information during the selection process. 

A viable way of deriving spectral index of a transient is still through multi-band flux measurements, and requiring the measurements be taken (at least quasi-) simultaneously. There are two ongoing surveys well-suited for this task -- VLASS and VCSS -- which scan the sky simultaneously at 3.0\,GHz and 340\,MHz, respectively. The order of magnitude separation in frequency provides significant leverage for characterizing the spectral slope of the radio continuum, and simultaneity ensures that the derived spectral indices are accurate for transient sources. Therefore, transient searches that combine VLASS and VCSS can incorporate spectral index and in theory, take advantage of the supplementary constraints encoded within this quantity.

In this study, we present our search for transients with inverted spectra ($\alpha > 0$) in epoch 1 of VLASS and VCSS. Considering the depth and cadence, these two surveys are excellent for finding bright, slow-evolving, extragalactic transients. Our primary objective is to determine whether it is feasible to discover and select interesting transients based on spectral index and to broadly evaluate the types of transients and physical constraints that result from such searches, particularly in VLASS and VCSS. In essence, this is a proof-of-concept study aimed to gauge expectations for transient selections that incorporate spectral information, which is a concept mostly unexplored in the current literature. Ideally, the results of our study can serve as inspiration for future transient searches as spectral information become more widely available from radio surveys (either from in-band spectrum or simultaneous multi-band measurements).

This study is structured as follows. In Section \ref{sec:methods}, we describe the radio surveys used and the method of our search for inverted-spectra transients. In Section \ref{sec:results}, we present our final sample of transient candidates along with our findings regarding multi-wavelength counterparts. In Section \ref{sec:discussion}, we discuss possible classifications for our transient candidates, examine the physical constraints associated with the inverted spectra, and reflect on the feasibility of our search method. Finally, we summarize and conclude our study in Section \ref{sec:conclusion}. Throughout this study, we assume a flat universe with $H_0 = 67.4\,\mathrm{km}\,\mathrm{s}^{-1}\,\mathrm{Mpc}^{-1}$ and $\Omega_\mathrm{m}=0.315$ \citep[][]{Plank2018VICosmoPar}.

\section{Methods} \label{sec:methods}

\begin{deluxetable*}{ccccccc}
\tablecaption{General Information on Relevant Radio Surveys\label{tab:surveys}}
\tablewidth{0pt}
\tablehead{
\colhead{Survey} & \colhead{Frequency} & \colhead{Sky Overlap} & \colhead{Epoch} & \colhead{Resolution} & \colhead{Image rms} & \colhead{Match Radius}\\
\nocolhead{} & \colhead{(MHz)} & \colhead{(deg$^2$)} & \nocolhead{} & \colhead{(arcsec)} & \colhead{(mJy/beam)} & \colhead{(arcsec)}
}
\startdata
VLASS Epoch 1 \& 2 & 3000 & \nodata & 2017--2022 & 2.5 & 0.14 & \nodata \\
VCSS  Epoch 1 \& 2 & 340 & 29800 & 2017--2022 & 20 & 3.0 & 5 \\
RACS-low & 888 & 23300  & 2019--2020 & 25 & 0.3 & 5 \\
RACS-mid & 1368 & 28900  & 2020--2022 & 10 & 0.2 & 3 \\
LoTSS DR2 & 144 & 5700 & 2014--2020 & 6 & 0.08 & 2 \\
NVSS & 1400 & 33880 & 1993--1996 & 45 & 0.45 & 5 \\
FIRST & 1400 & 10800 & 1993--2004, 2009-2011 & 5 & 0.15 & 3 \\
WENSS & 325 &  10000 & 1990s & 54 & 3.6 & 5 \\
TGSS & 150 & 31500 & 2010--2012 & 25 & 3.5 & 5 \\
AT20G & 4800, 8640, 20000 & 13000 & 2004-2008 & 10 & 10 & 4
\enddata
\tablecomments{The sky overlap is the survey area overlapping with VLASS crudely estimated using HEALPix \citep[][]{Gorski2005HEALPix}. Note that although some surveys are ongoing, we only list the epochs of the data used in this study. The match radius is for cross-matching with the VLASS catalogue, chosen based on the astrometric uncertainty of the survey. See Section~\ref{subsec:surveydescription} for references.}
\end{deluxetable*}

\subsection{Description of Radio Surveys} \label{subsec:surveydescription}

VLASS \citep[][]{Lacy2020VLASS} is a recent survey with the VLA that began in September 2017. VLASS covers the entire sky North of $-40^{\circ}$ declination ($\sim$82\% or 33,885\,deg$^2$ of the sky) in the 3.0\,GHz (2.0--4.0\,GHz) band over each observing epoch with an angular resolution of 2.5\,arcsec. Originally, VLASS was designed to have three observing epochs, each separated by $\sim$1000\,d, to be completed in October 2024. At the time of writing, a fourth half-epoch was approved for VLASS that will be conducted over 2025--2026. For this study, we utilize the VLASS component catalogs produced by the Canadian Initiative for Radio Astronomy Data Analysis\footnote{See \url{https://cirada.ca/vlasscatalogueql0}.} \citep[CIRADA;][]{Gordon2021VLASS}{}{} using VLASS quick look (QL) images\footnote{QL images are rapidly-processed images generated from a relatively simple imaging algorithm with some known quality issues in flux density and astrometry \citep[][]{Memo13}{}{}.}. Specifically, we use the epoch 1 catalog with 3,347,423 components (version 3) for the transient search and we also use the epoch 2 catalog (version 2) to examine variability. The median rms of the QL images is $\sim$140$\,\mu\mathrm{Jy/beam}$.

VCSS \citep[][]{Peters2021VCSS} is a survey conducted simultaneously with VLASS by VLITE, a commensal instrument on the VLA developed jointly by the U.S. Naval Research Laboratory and the NRAO \citep[][]{Clarke2016VLITE,Polisensky2016,Polisensky2024}. VCSS images are processed through the VLITE Database Pipeline \citep[][]{Polisensky2019VDP}. VCSS covers the same regions of the sky as VLASS in the 340\,MHz (320--384\,MHz) band (although with a wider field of view at lower frequency) with an angular resolution of 20\,arcsec. However, we note that the VCSS sky coverage has some empty regions or holes because VLITE was offline or imaging was unfeasible due to insufficient data or nearby bright sources \citep[see][]{VCSSBrightmemo}. For the transient search, we utilize the VCSS epoch 1 catalog with 573,293 components generated using the Python Blob Detector and Source Finder \citep[PyBDSF;][]{MohanRafferty2015PyBDSF}. The median rms of the VCSS epoch 1 images is $\sim$3\,mJy/beam.

To search for transients in VLASS and VCSS, we make use of available catalogs and images from past radio surveys, including the NRAO VLA Sky Survey \citep[NVSS;][]{Condon1998NVSS}, the Faint Images of the Radio Sky at Twenty-Centimeters \citep[FIRST, 14Dec17 version;][]{Becker1995FIRST,Helfand2015FIRST14Dec17} survey, the Westerbork Northern Sky Survey \citep[WENSS;][]{Rengelink1997WENSS}, the TIFR Giant Metrewave Radio Telescope Sky Survey Alternative Data Release \citep[TGSS;][]{TGSS}, and the Australia Telescope 20 GHz \citep[AT20G;][]{AT20G} survey. We use all these past surveys as templates for the transient search (Section~\ref{subsec:searchmethod}).

Lastly, after obtaining a list of transient candidates, we also search three recent surveys, RACS data release (DR) 1 in the lowest band \citep[RACS-low;][]{McConnell2020RACSI,Hale2021RACSIIcatalog}, RACS in the mid band \citep[RACS-mid][]{Duchesne2023RACSmid,Duchesne2023RACSmidCatalog}, and LoTSS DR2 \citep[][]{Shimwell2022LoTSSDR2}. These surveys were carried out at times near epoch 1 and 2 of VLASS but at lower frequencies, which we utilize to confirm detections and obtain further spectral coverage. Table \ref{tab:surveys} outlines some general information regarding the radio surveys described above.

\subsection{Search for Inverted-Spectra Transients} \label{subsec:searchmethod}

The goal of our search is to obtain a list of transient candidates in epoch 1 of VLASS and VCSS with inverted spectra. Our general strategy involves identifying transient candidates through comparing VLASS with past radio surveys, and identifying candidates with inverted spectra through comparing VLASS with VCSS. We use a number of selection criteria to ensure that our final list of candidates is robust. The steps of our search are described below and the number of remaining sources after each step is listed in Table~\ref{tab:numberaftersteps}.

\bigskip \noindent{\underline{\textbf{1. Reduce Contamination in VLASS:}}}
\medskip

Image artifacts can easily be misidentified as transients. To minimize contamination from artifacts, we follow the recommendations of the CIRADA catalog User Guide (July 26, 2023 version)\footnote{\url{https://ws.cadc-ccda.hia-iha.nrc-cnrc.gc.ca/files/vault/cirada/tutorials/CIRADA__VLASS_catalogue_documentation_2023_june.pdf}} and select components from the VLASS epoch 1 catalog with $Quality\_flag=0|4|8|12$, $P\_sidelobe<0.05$, $S\_Code\neq\mathrm{E}$, and $Duplicate\_flag<2$. The resulting sample from this selection has an estimated contamination (i.e., number of real sources over number of artifacts) of only $\sim$0.12\%. This sample is then used in the following steps.

\bigskip \noindent{\underline{\textbf{2. Compact VLASS Sources:}}}
\medskip

Considering the cadence of VLASS ($\sim$1000\,days) and the brightness threshold of inverted-spectra sources (lower limit set by VCSS $5\sigma\sim15\,\mathrm{mJy}$; see step 3 below), we expect to mainly find bright extragalactic synchrotron transients \citep[][]{Pietka2015}. Based on the light travel time argument, sources at extragalactic distances that vary in brightness over multiple years should be compact in VLASS\footnote{A source that is variable over $t\sim10\,\mathrm{yr}$ with emission size $ct$ will only be resolved if it is at a distance of $\lesssim$250\,kpc given the resolution of VLASS of 2.5\,arcsec.}. Therefore, we first identify compact VLASS sources for the transient search.

\begin{figure}
\epsscale{1.14}
\plotone{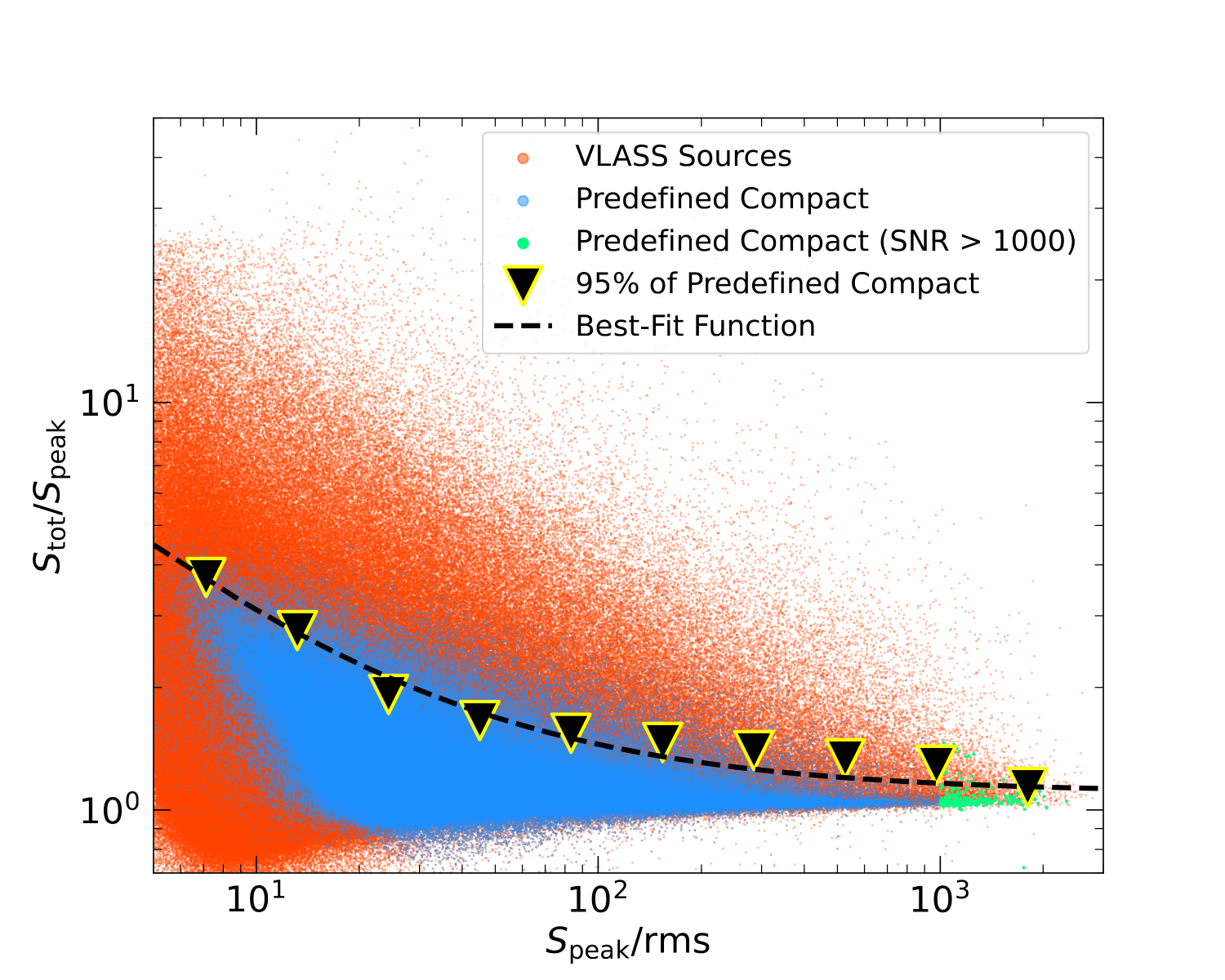}
\caption{Total flux density over peak flux density versus peak flux density over local noise for VLASS sources (red points). Blue points represent the predefined compact sample and green points represent those with $\mathrm{SNR} > 1000$. The black triangles enclose 95\% of the blue points within each bin. The dashed line is the best-fit envelope from fitting the black triangles. All points below the dashed line are defined to be compact VLASS sources. \label{fig:VLASSCompact}}
\end{figure}

We define a compactness criterion following the method described by \citet{deRuiter2021} and \citet{Shimwell2019LoTSSDR1} to identify compact sources. First, we construct a parameter space, shown in Figure \ref{fig:VLASSCompact}, consisting of the ratio of total flux density to peak flux density ($\mathrm{S}_{\mathrm{tot}}/\mathrm{S}_{\mathrm{peak}}$) and the ratio of peak flux density to local noise ($\mathrm{S}_{\mathrm{peak}}/\mathrm{rms}$). We then select a sample of predefined compact sources in VLASS that satisfy the following conditions: (i) $S\_Code=\mathrm{S}$, i.e., well fit by a single Gaussian, (ii) total flux density greater than 3\,mJy to ensure completeness \citep[][]{Gordon2021VLASS}, (iii) nearest neighboring source more than 45\,arcsec away (twice the combined beam size of VLASS and VCSS), and (iv) component major axis smaller than 5\,arcsec (twice the nominal VLASS resolution). Note that we adopt 5\,arcsec in criterion (iv) to account for non-uniform beam sizes at very high/low declinations or variations due to imperfect calibration. While there is no definitive value for this threshold, we chose a conservative value to retain as many compact sources as possible and exclude the most extended sources. For reference, this criterion excludes less than 1\% of the components with a deconvolved major axis size of 0\,arcsec while excluding $\sim$13\% of the entire sample at this stage.

We divide the predefined compact sources into ten bins over $\mathrm{S}_{\mathrm{peak}}/\mathrm{rms}$, and for each bin, we find the $\mathrm{S}_{\mathrm{tot}}/\mathrm{S}_{\mathrm{peak}}$ value that is above 95\% of the predefined compact sources in the bin. Finally, we fit the 95\% levels with the function
\begin{equation}
    \frac{\mathrm{S}_{\mathrm{tot}}}{\mathrm{S}_{\mathrm{peak}}} = \mathrm{offset} + \mathrm{A}\cdot \left( \frac{\mathrm{S}_{\mathrm{peak}}}{\mathrm{rms}} \right)^{\mathrm{B}},
\end{equation}
where the offset is defined to be the median plus three times the median absolute deviation of $\mathrm{S}_{\mathrm{tot}}/\mathrm{S}_{\mathrm{peak}}$ of the predefined compact sources with $\mathrm{S}_{\mathrm{peak}}/\mathrm{rms}>1000$. We find the best-fit parameters to be $\mathrm{A}=11.48$, $\mathrm{B}=-0.76$, and $\mathrm{offset}=1.10$. We define all VLASS sources below this function in the parameter space to be compact sources (i.e., all points below the black dashed line in Figure \ref{fig:VLASSCompact}).

\bigskip \noindent{\underline{\textbf{3. Inverted Spectra in VLASS and VCSS}}}
\medskip

In this step, we cross-match the compact VLASS epoch 1 sources with the VCSS epoch 1 catalog to identify compact sources with inverted spectra. Considering the astrometric uncertainty of VLASS and VCSS, we use a match radius of 5\,arcsec. Additionally, to exclude empty regions of the VCSS sky coverage from our search, we remove VLASS sources that have no neighboring VCSS sources within 15\,arcmin. From matching, we obtain two populations: (i) sources detected by both VLASS and VCSS and (ii) sources detected by VLASS but not VCSS. Note that we do not consider sources detected by VCSS but not VLASS because these would not have inverted spectra.

Before estimating spectral index, we apply flux density corrections to the catalogs, as both VLASS QL and VCSS catalogs are known to have underestimated flux densities \citep[][]{Memo13,VCSSBrightmemo}{}{}. For the VLASS epoch 1 catalog, following the analysis in Appendix \ref{apdx:FluxCal}, for epoch 1.1 and 1.2, we apply a correction factor of 1/0.95 and 1/0.98 to the total flux densities and add an additional 7\% and 3\% of the total flux densities in quadrature to the uncertainty, respectively. For the VCSS epoch 1 catalog, we apply the correction function\footnote{The correction is in the form of $S_{\mathrm{obs}}/S_{\mathrm{corr}} = ae^{-b\cdot\mathrm{SNR}}+c$ where $a=-0.382$, $b=0.032$, $c=0.899$ for total flux density and SNR is the signal-to-noise ratio.} shown in Figure 2 of \citet{VCSSBrightmemo}.

We use the corrected total flux densities to estimate spectral indices. For sources detected in both VLASS and VCSS, we directly calculate their spectral indices between 340\,MHz and 3.0\,GHz\footnote{Note that we assume the VLASS flux density to be at 3.0\,GHz but technically, it is over a band of 2.0--4.0\,GHz with a large fractional bandwidth. This introduces a small error of less than 5\% in the flux density for a power law spectrum with $\alpha$ between -1.0 and 2.0, which should not significantly impact our analysis.} assuming $S_\nu \propto \nu^\alpha$, where $S_\nu$ is the total flux density and $\alpha$ is the spectral index. Errors of the indices, $\sigma_{\alpha}$, are estimated from propagating the flux density errors. For sources detected in VLASS but not VCSS, we assume an upper limit of $S_{340\mathrm{MHz}} < 15\,\mathrm{mJy}$, corresponding to the average $5\sigma$ level of VCSS, and estimate a lower limit on $\alpha$. We adopt a conservative $5\sigma$ value for this search to account for the non-uniform sensitivity of VCSS \citep[][]{VCSSBrightmemo}, as we do not attempt to derive local upper limits (or forced photometry) at the location of every VLASS source.

Finally, we identify sources with inverted spectra by selecting those with $\alpha > 0$. For sources detected in both VLASS and VCSS, we select those with $\alpha - 3\sigma_{\alpha} > 0$. For sources detected in VLASS but not VCSS, we select those with $S_{3.0\,\mathrm{GHz}} - 3\sigma_{S_{3.0\,\mathrm{GHz}}} > 0$, where $\sigma_{S_{3.0\,\mathrm{GHz}}}$ is the error of the flux density in VLASS. Note that limited by the sensitivity of VCSS, $S_{3.0\,\mathrm{GHz}} \gtrsim 15\,\mathrm{mJy}$ is required for $\alpha > 0$, which implies a detection level of $\gtrsim$100$\sigma$ in VLASS. Therefore, VLASS sources selected from this step are highly unlikely to be random noise structures.

\begin{deluxetable*}{lccc}
\tablecaption{Number of VLASS Sources Remaining After Each Step of Transient Search\label{tab:numberaftersteps}}
\tablewidth{0pt}
\tablehead{
\colhead{Step} & \colhead{Remaining VLASS Sources} & \colhead{VCSS Detection} & \colhead{VCSS Non-detection}
}
\startdata
1. Reduce Contamination & 1,660,228 & \nodata & \nodata  \\
2. Compact Sources & 1,440,639 & \nodata & \nodata   \\
3. Inverted Spectra & 14,628 & 1,320 & 13,308 \\
4. Potential Transients & 1,460 & 1 & 1,459 \\
5. Removing AGNs & 1,190 & 1 & 1,189 \\
6. Visual Inspection & 21 & 1 & 20
\enddata
\tablecomments{The last two columns show the number of VLASS sources detected and not detected in VCSS.}
\end{deluxetable*}

\bigskip \noindent{\underline{\textbf{4. Identifying Potential Transients}}}
\medskip

After selecting compact VLASS sources with inverted spectra, we then identify the transient candidates in this sample. For this study, we consider a transient candidate to be a source that was detected in VLASS epoch 1 but was not detected in any of the past radio surveys listed in Table \ref{tab:surveys}. This definition of a transient is pragmatic and is designed to capture sources displaying significant variability likely associated with transient phenomena. 

In this step, we cross-match the compact inverted-spectra VLASS sources with the NVSS, FIRST, TGSS, WENSS, and AT20G catalogs (using the match radii listed in Table \ref{tab:surveys} chosen based on astrometric uncertainties of the surveys) and exclude all sources with a match. Remaining sources without a match would then be potential inverted-spectra transients.

We note that for this step, the first and most important comparison is with NVSS, which has the closest frequency (1.4\,GHz), identical sky coverage, and adequate depth (typical $5\sigma$ upper limit of $\sim$2.3\,mJy). Other past surveys, which partially overlap with VLASS or cover different frequencies, are compared after NVSS for additional confirmation. For reference, the inclusion of FIRST, WENSS, TGSS, and AT20G removes an additional 946 sources after NVSS (from 2406 to 1460 sources). We also visually inspected a number of these removed sources and found that they were not cross-matched to NVSS but were cross-matched to other surveys because of the same issues discussed in step 6 below (e.g., resolution and astrometric uncertainty of NVSS).

\bigskip \noindent{\underline{\textbf{5. Removing Known Active Galactic Nuclei}}}
\medskip

AGNs are known to manifest as variable radio sources and are not considered traditionally as transient sources \citep[][]{Barvainis2005}. Thus, variable AGNs are not a target of interest in this study. In an attempt to reduce AGN ``contamination'', we remove known/cataloged AGNs by cross-matching our sample with the Wide-field Infrared Survey Explorer (WISE) R90 AGN catalog \citep[][]{Assef2018WISEAGN}, the Sloan Digital Sky Survey (SDSS) Quasar catalog \citep[DR16Q\_v4;][]{Lyke2020SDSSQSODR16}, and the Milliquas catalog \citep[v8;][]{Milliquasv64,Milliquasv72,Milliquasv8}. We exclude all sources with a match within a radius of 3.3\,arcsec following the finding of \citet{DAbrusco2013}. This step should be effective in removing bright AGNs that dominate the emission at optical and infrared (IR) wavelengths. However, it does not exclude AGNs with faint optical and IR emission or AGNs too distant to be reliably classified. Therefore, in the following sections, we will still consider the possibility that our final sample may contain AGNs, particularly those that are presumably weaker or heavily absorbed and have then brightened significantly in the radio (see Section~\ref{subsubsec:AGNVar}).

\bigskip \noindent{\underline{\textbf{6. Visual Inspection}}}
\medskip

In this last step, we visually inspect the survey images of the remaining sources for a final check and reject sources that we consider to be false positives (such as artifacts or persistent sources) misidentified as transients. Generally, we find false positives to be caused by (i) resolution difference between surveys, (ii) outliers with relatively large astrometric uncertainty, or (iii) degraded local image quality in past surveys.

Due to the resolution difference between VLASS and NVSS (2.5\,arcsec versus 45\,arcsec), we often find false positives when multiple neighboring VLASS sources are blended into a single NVSS source. In this case, the centroid of the blended NVSS source can be significantly offset from the VLASS position, meaning that the VLASS source will not be matched to the NVSS source within our chosen radius of 5\,arcsec. As a result, the VLASS source is misidentified as a potential transient, which we reject as a false positive. Also, we find that this issue exists for all remaining sources that have neighboring VLASS sources within 45\,arcsec (resolution of NVSS). For future searches, it may be possible to reduce the impact of resolution difference by introducing an additional step of rejecting sources that are not isolated (e.g., reject sources with a neighbor within the beam size of the lower resolution survey).

Occasionally, for isolated VLASS sources, we still see a nearby NVSS source at $\sim$5--10\,arcsec, slightly farther away than our chosen match radius of 5\,arcsec. While in theory, it is possible that these are two transients -- a disappearing NVSS source and an appearing VLASS source -- coincidentally nearby, a more likely explanation is that the two sources are associated but have relatively large astrometric uncertainties, especially considering the scattering in the astrometry of VLASS \citep[Figure 8 of][]{Gordon2021VLASS} and NVSS \citep[Figure 28 and 29 of][]{Condon1998NVSS}. Therefore, to ensure that our final candidates are robust, we also reject these sources as false positives.

\begin{figure*}
\epsscale{1.14}
\plotone{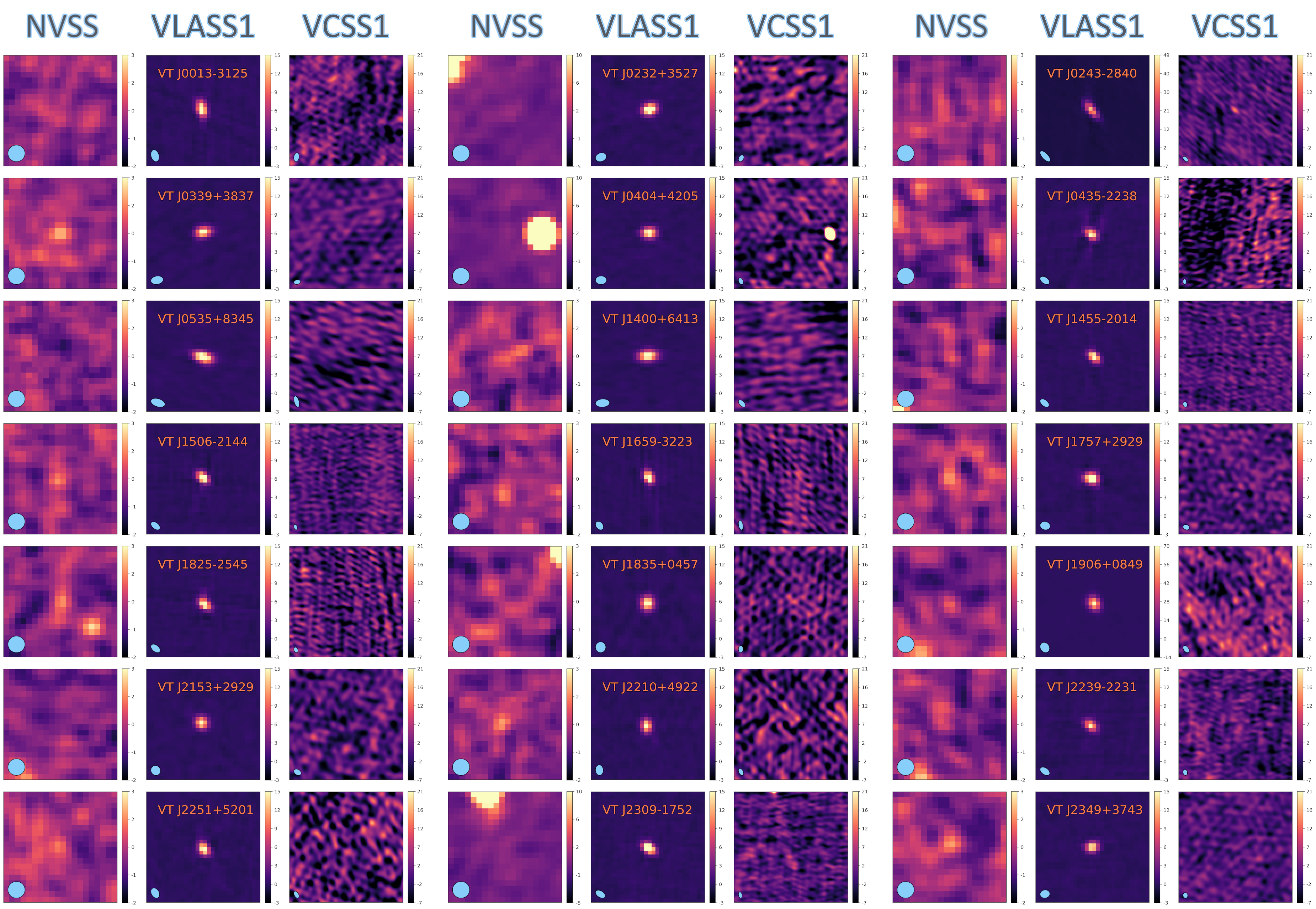}
\caption{Radio cutout images centered at the locations of the inverted-spectra transient candidates from NVSS, VLASS epoch 1 (VLASS1), and VCSS epoch 1 (VCSS1), arranged into three columns. The names of the candidates are shown in the VLASS1 images. The image size (side length) is 5\,arcmin, 30\,arcsec, and 5\,arcmin for NVSS, VLASS1, and VCSS1, respectively. For each image, the synthesized radio beam (blue ellipse) is shown at the lower left and a colorbar (in mJy) is shown on the right side. \label{fig:RadioImages}}
\end{figure*}

We typically find degraded NVSS image quality in the Galactic plane near extremely bright emission, where local flux density can fluctuate by tens of mJy over large spatial scales. In this case, a non-detection in NVSS can only provide a shallow limit, often exceeding the measured VLASS flux density of $\gtrsim$15\,mJy, which cannot be used as evidence of significant variability between NVSS and VLASS. Therefore, we reject sources in regions with degraded NVSS image quality as false positives.

In addition, we rejected one case (at RA=222.32510\,deg and Dec=76.01997\,deg) that was only detected in VLASS epoch 1.1 (appearing as a transient). It has a signal-to-noise ratio of $\gtrsim$60 and does not appear to be a \textsc{clean} artifact. However, this source is 178\,arcsec away from a bright source in the right ascension direction, which matches the description of ghost artifacts found in VLASS epoch 1.1 that were caused by delay center errors \citep[][]{Memo13}. Therefore, based on the suspicion that it could be a VLASS ghost artifact, we excluded it from our final sample.

After rejecting the false positives described above, the remaining sources are considered to be real transient candidates. Ultimately, we found 21 inverted-spectra transient candidates from our search procedure. Figure \ref{fig:RadioImages} shows cutout images of the candidates from NVSS, VLASS epoch 1, and VCSS epoch 1. We also show cutout images from the other radio surveys in Figure~\ref{fig:OtherRadioImg1} and Figure~\ref{fig:OtherRadioImg2} in Appendix~\ref{apdx:OtherRadioImg}.

\section{Results} \label{sec:results}

Our search resulted in a sample of 21 transient candidates with inverted spectra (between 340\,MHz and 3.0\,GHz) in epoch 1 of VLASS and VCSS, with one detected and 20 not detected in VCSS. After the search, for temporal coverage, we obtained measurements for the candidates from epoch 2 of VLASS and VCSS (note that some candidates lack VCSS epoch 2 measurements since they fall on the empty regions of the VCSS epoch 2 sky coverage). For more comparison and spectral coverage, we also searched RACS-low, RACS-mid, and LoTSS DR2 using match radii listed in Table~\ref{tab:surveys}. However, we do not use RACS or LoTSS for deriving spectral indices because their observations are not taken at the same time as VLASS and VCSS.

Table \ref{tab:candobstable} contains the collected radio properties of our transient candidates, including names, positions, observing times, radio flux densities, and VLASS-VCSS spectral indices. We follow previous studies and adopt the letters VT (VLASS Transient) for names. Note that the tabulated flux density upper limits are $3\sigma$ limits where $\sigma$ is the local rms calculated using a box centered at the locations of the candidates. The box size (side length) is chosen to be 120\,arcsec, 120\,arcsec, 30\,arcsec, 150\,arcsec, 90\,arcsec, and 90\,arcsec for NVSS, VCSS, VLASS, RACS-low, RACS-mid, and LoTSS DR2, respectively, to capture local noise and avoid nearby sources. As seen in the images (Figure~\ref{fig:RadioImages}), the backgrounds are generally uniform and Gaussian, meaning that our derived $3\sigma$ values should be accurate representations of the local flux density upper limits. The tabulated lower limits of $\alpha$ are also calculated using the local $3\sigma$ flux density limits, which are all positive as expected. We also show optical cutout images at the locations of the candidates in Figure \ref{fig:OpticalImages} (details discussed in Section \ref{subsubsec:galaxysearch} and \ref{subsubsec:opticalIRsearch}). 

At the time of writing, only three of our transient candidates -- VT\,J0243-2840, VT\,J0535+8345, and VT\,J2239-2231 -- have been reported by recent searches of ongoing radio surveys in the literature to the best of our knowledge. The discovery of VT\,J0243-2840 was first reported by \citet{Somalwar2023J0243}, which resulted from two independent searches: one search compared VLASS and NVSS \citep[similar to the procedure described by][]{Dong2023} and the other was our search for inverted-spectra transients (i.e., this study). \citet{Kunert-Bajraszewska2025} reported all three of these candidates from a search for nuclear transients between NVSS and VLASS. We are also aware that VT\,J1906+0849 has been discovered separately with extensive follow-up observations that will be reported in a future study (J. Miller et al., in prep.). Nonetheless, we appear to have found a new sample of transient candidates, which already hints the potential of transient discovery in radio surveys and inverted spectra as a selection criteria. In this section, we examine the properties of our transient candidates in detail. In particular, we discuss our search for multi-wavelength counterparts (Section \ref{subsec:searchmultiwavecounterpart}) and the general radio properties of the candidates (Section \ref{subsec:generalradioproperty}).

\begin{splitdeluxetable*}{cccccccccccBccccccccc}
\tablecaption{Radio Measurements of the VLASS-VCSS Epoch 1 Inverted-Spectra Transient Candidates\label{tab:candobstable}}
\tablewidth{0pt}
\tabletypesize{\scriptsize}
\tablehead{
\colhead{Name} & \colhead{R.A.} & \colhead{Decl.} & \colhead{MJD} & 
\colhead{S$_{1.4\mathrm{GHz}}$} & \colhead{MJD} & 
\colhead{S$_{3.0\mathrm{GHz}}$} & \colhead{S$_{340\mathrm{MHz}}$} & 
\colhead{$\alpha$} & \colhead{MJD} & \colhead{S$_{3.0\mathrm{GHz}}$} & \colhead{S$_{340\mathrm{MHz}}$} & 
\colhead{$\alpha$} & \colhead{MJD} & \colhead{S$_{888\mathrm{MHz}}$} & \colhead{MJD} & \colhead{S$_{1368\mathrm{MHz}}$} & \colhead{MJD} & \colhead{S$_{150\mathrm{MHz}}$} & \colhead{Spectrum} \\
\colhead{} & \colhead{(J2000)} & \colhead{(J2000)} & \colhead{(NVSS)} & \colhead{(NVSS)} & \colhead{(Epoch 1)} & \colhead{(VLASS1)} &
\colhead{(VCSS1)} & \colhead{(Epoch 1)} & \colhead{(Epoch 2)} & \colhead{(VLASS2)} & \colhead{(VCSS2)} & \colhead{(Epoch 2)} & \colhead{(RACS-low)} & \colhead{(RACS-low)} & \colhead{(RACS-mid)} & \colhead{(RACS-mid)} & \colhead{(LoTSS DR2)} & \colhead{(LoTSS DR2)} \\
\colhead{} & \colhead{(deg)} & \colhead{(deg)} & \colhead{(d)} & 
\colhead{(mJy)} & \colhead{(d)} & \colhead{(mJy)} &
\colhead{(mJy)} & \colhead{} & \colhead{(d)} & \colhead{(mJy)} & \colhead{(mJy)} & \colhead{} & \colhead{(d)} & \colhead{(mJy)} & \colhead{(d)} & \colhead{(mJy)} & \colhead{(d)} & \colhead{(mJy)} & \colhead{}
}
\startdata
VT\,J0013-3125 & 3.27815   & -31.41743 & 49269 & $<$1.3 & 58665.5 & 20.5$\pm$0.7 & $<$9.5            & $>$0.35         & 59625.9 & 19.8$\pm$0.6 & $<$15.0           & $>$0.13         & 58936.1 & $<$1.0            & 59220.4 & 2.0$\pm$0.3  & \nodata       & \nodata    &   Highly Inverted               \\
VT\,J0232+3527 & 38.07406  & 35.46137  & 49336 & $<$1.7 & 58645.5 & 22.6$\pm$0.7 & $<$11.5           & $>$0.31         & 59523.4 & 25.6$\pm$0.8 & $<$13.4           & $>$0.30         & 58594.3 & 2.3$\pm$0.7  & 59207.5 & 8.1$\pm$0.9  & 58420.8 & 3.1$\pm$0.2 & Highly Inverted$^*$ \\
VT\,J0243-2840 & 40.94038  & -28.67775 & 49250 & $<$1.6 & 58166.1 & 53.8$\pm$3.6 & 31.4$\pm$4.5 & 0.25$\pm$0.07 & 59154.2 & 59.2$\pm$1.8 & 19.2$\pm$6.3 & 0.52$\pm$0.15 & 58601.2 & 45.8$\pm$3.7 & 59210.5 & 65.1$\pm$3.9 & \nodata       & \nodata                 &     Peaked  \\
VT\,J0339+3837 & 54.84758  & 38.63292  & 49336 & $<$2.6 & 58586.7 & 17.8$\pm$0.6 & $<$7.2            & $>$0.42         & 59546.3 & 14.4$\pm$0.5 & $<$7.4            & $>$0.30         & 58594.3 & 15.2$\pm$1.6 & 59207.5 & 13.4$\pm$0.9 & \nodata       & \nodata                  &   Peaked   \\
VT\,J0404+4205 & 61.22167  & 42.08469  & 49336 & $<$3.6 & 58617.7 & 18.8$\pm$0.6 & $<$11.8           & $>$0.21         & 59511.5 & 15.7$\pm$0.5 & $<$12.1           & $>$0.12         & \nodata       & \nodata                         & 59206.6 & 4.5$\pm$0.4  & \nodata       & \nodata                    &   Highly Inverted \\
VT\,J0435-2238 & 68.86741  & -22.64586 & 49250 & $<$1.8 & 58161.0 & 19.2$\pm$1.3 & $<$17.2           & $>$0.05         & 59149.4 & 19.8$\pm$0.7 & $<$9.5            & $>$0.34         & 58599.5 & $<$1.0            & 59237.6 & 2.7$\pm$0.3  & \nodata       & \nodata                    &  Highly Inverted  \\
VT\,J0535+8345 & 83.79209  & 83.76150  & 49280 & $<$1.5 & 58074.2 & 23.7$\pm$1.6 & $<$11.5           & $>$0.33         & 59104.3 & 27.4$\pm$0.8 & $<$7.7            & $>$0.58         & \nodata       & \nodata                         & \nodata       & \nodata                         & \nodata       & \nodata                 &   \nodata    \\
VT\,J1400+6413 & 210.00266 & 64.21716  & 49314 & $<$2.0 & 58084.5 & 20.4$\pm$1.4 & $<$11.1           & $>$0.28         & 59080.0 & 19.0$\pm$0.6 & $<$8.7            & $>$0.36         & \nodata       & \nodata                         & \nodata       & \nodata                         & 58773.3 & 0.8$\pm$0.2 & Highly Inverted \\
VT\,J1455-2014 & 223.77882 & -20.24222 & 50238 & $<$1.5 & 58162.5 & 22.0$\pm$1.5 & $<$6.4            & $>$0.57         & 59147.8 & 26.0$\pm$0.9 & $<$8.4            & $>$0.52         & 58600.6 & $<$0.6            & 59424.4 & 2.0$\pm$0.3  & \nodata       & \nodata                    &   Highly Inverted \\
VT\,J1506-2144 & 226.68930 & -21.74295 & 50238 & $<$1.9 & 58162.5 & 25.1$\pm$1.7 & $<$7.5            & $>$0.55         & 59147.8 & 14.9$\pm$0.5 & $<$9.0            & $>$0.23         & 58934.8 & $<$0.8            & 59276.8 & 3.2$\pm$0.4  & \nodata       & \nodata                    &  Highly Inverted  \\
VT\,J1659-3223 & 254.86653 & -32.39264 & 49735 & $<$1.8 & 58153.6 & 22.9$\pm$1.6 & $<$12.3           & $>$0.29         & 59173.8 & 19.9$\pm$0.7 & $<$9.1            & $>$0.36         & 58601.9 & 3.0$\pm$0.9  & 59231.1 & 7.2$\pm$0.5  & \nodata       & \nodata                    &   Highly Inverted \\
VT\,J1757+2929 & 269.38016 & 29.49229  & 49823 & $<$2.3 & 58028.1 & 26.0$\pm$1.7 & $<$7.9            & $>$0.54         & 59103.2 & 30.5$\pm$1.0 & $<$8.6            & $>$0.58         & 58594.8 & $<$0.9            & 59218.1 & 3.5$\pm$0.4  & 58697.7 & 1.4$\pm$0.4 & Highly Inverted$^*$ \\
VT\,J1825-2545 & 276.45508 & -25.75572 & 50238 & $<$2.3 & 58166.6 & 24.9$\pm$1.7 & $<$10.8           & $>$0.38         & 59146.9 & 24.2$\pm$0.9 & $<$9.5            & $>$0.43         & 59020.6 & 4.2$\pm$1.0  & 59231.1 & 8.8$\pm$0.6  & \nodata       & \nodata                     &  Highly Inverted \\
VT\,J1835+0457 & 278.91964 & 4.95932   & 49775 & $<$1.7 & 58559.6 & 17.9$\pm$0.6 & $<$11.8           & $>$0.19         & 59485.1 & 20.2$\pm$0.7 & $<$10.3           & $>$0.31         & 58971.8 & $<$0.8            & 59238.1 & $<$0.6            & \nodata       & \nodata          &             Highly Inverted  \\
VT\,J1906+0849 & 286.57567 & 8.82469   & 50268 & $<$1.7 & 58053.1 & 76.6$\pm$5.1 & $<$13.1           & $>$0.81         & 59092.0 & 37.5$\pm$1.3 & \nodata                         & \nodata                          & 58972.9 & 52.3$\pm$4.4 & 59236.2 & 40.9$\pm$2.5 & \nodata       & \nodata                  &  Peaked    \\
VT\,J2153+2929 & 328.49995 & 29.49528  & 49853 & $<$1.2 & 58625.4 & 17.6$\pm$0.6 & $<$9.4            & $>$0.29         & 59489.3 & 21.4$\pm$0.7 & $<$12.8           & $>$0.24         & 58598.0 & $<$0.9            & 59216.3 & 2.3$\pm$0.4  & 58758.7 & $<$0.5      &   Highly Inverted  \\
VT\,J2210+4922 & 332.57300 & 49.37413  & 49793 & $<$2.1 & 58626.5 & 17.1$\pm$0.6 & $<$15.2           & $>$0.06         & 59505.3 & 13.8$\pm$0.5 & $<$7.6            & $>$0.27         & \nodata       & \nodata                         & \nodata       & \nodata                         & \nodata       & \nodata                  &    \nodata  \\
VT\,J2239-2231 & 339.88860 & -22.52422 & 49250 & $<$1.9 & 58665.4 & 17.3$\pm$0.6 & $<$7.1            & $>$0.41         & 59625.8 & 15.5$\pm$0.6 & \nodata                         & \nodata                          & 58601.1 & 9.1$\pm$1.3  & 59208.4 & 16.5$\pm$1.0 & \nodata       & \nodata                  &    Peaked  \\
VT\,J2251+5201 & 342.92041 & 52.03047  & 49788 & $<$2.4 & 58626.5 & 19.3$\pm$0.6 & $<$13.2           & $>$0.18         & 59514.3 & 18.4$\pm$0.6 & $<$11.1           & $>$0.23         & \nodata       & \nodata                         & \nodata       & \nodata                         & \nodata       & \nodata                 &     \nodata  \\
VT\,J2309-1752 & 347.42478 & -17.88210 & 49250 & $<$2.2 & 58670.5 & 27.3$\pm$0.9 & $<$9.1            & $>$0.51         & 59614.8 & 18.1$\pm$0.7 & \nodata                         & \nodata                          & 58601.1 & 11.1$\pm$1.4 & 59235.3 & 16.4$\pm$1.0 & \nodata       & \nodata                  &    Peaked  \\
VT\,J2349+3743 & 357.33369 & 37.72596  & 49835 & $<$2.3 & 58593.6 & 16.9$\pm$0.5 & $<$5.8            & $>$0.50         & 59510.3 & 14.5$\pm$0.5 & $<$5.5            & $>$0.44         & 58595.1 & 3.1$\pm$0.7  & 59209.4 & 8.9$\pm$0.7  & \nodata       & \nodata                 &    Highly Inverted \\
\enddata
\tablecomments{The symbol $<$ and $>$ indicates upper limit and lower limit, respectively. Note that flux density upper limits are $3\sigma$ where $\sigma$ is the local rms. Values after the $\pm$ symbol are $1\sigma$ errors. The spectral index $\alpha$ is calculated using only S$_{3.0\mathrm{GHz}}$ and S$_{340\mathrm{MHz}}$. The final column indicates the rough spectral shape if measurements at different frequencies are connected, and $*$ is marked if additional complexity is present in the spectrum (see Section~\ref{subsubsec:radioSEDs} for details).}
\end{splitdeluxetable*}

\subsection{Multi-wavelength Counterpart and Classification}\label{subsec:searchmultiwavecounterpart}

\subsubsection{Searching for Archival Transient Counterparts}\label{subsec:archivetransient}

For our inverted-spectra transient candidates, we searched the Transient Name Server\footnote{\url{https://www.wis-tns.org/}}, the Open Supernova Catalog\footnote{\url{https://sne.space/}} \citep[][]{Guillochon2017OSC}, and the Open TDE Catalog\footnote{\url{https://tde.space/}} for potential archival transient counterparts, particularly SNe and TDEs, before or after the VLASS observations. We found no matches within 1\,arcmin of our candidates. 

We also searched the GRBweb\footnote{\url{https://user-web.icecube.wisc.edu/~grbweb_public/}}, which contains GRBs retrieved from GCN-circulars\footnote{\url{https://gcn.nasa.gov/circulars}}, Fermi Gamma-ray Burst Monitor catalog \citep[][]{vonKienlin2020FermiGBMIV}, Fermi Large Area Telescope \citep[][]{Atwood2009FermiLAT}, Swift \citep[][]{Gehrels2004Swift}, interplanetary network\footnote{\url{https://www.mpe.mpg.de/~jcg/grbgen.html}}, BeppoSAX catalog \citep[][]{Boella1997BeppoSAX}, and the Burst And Transient Source Experiment catalog \citep[][]{Paciesas1999BATSE}. Against well-localized GRBs with a $1\sigma$ position error of $\lesssim$1\,arcsec, we found matches to be at least separated by 0.4\,deg ($\gtrsim$1000$\sigma$). Against GRBs with $1\sigma$ position error $<$1\,deg, we found no significant matches, as all but one of our transient candidates are outside the error circles by at least $3\sigma$. Only VT\,J0232+3527 is within the $2.8\sigma$ error circle of GRB\,920721D that has a fairly large $1\sigma$ position error of 0.73\,deg. Against GRBs with poorer localizations ($1\sigma$ position error $\geq$1\,deg and $<$10\,deg), we found that every one of our transient candidates are matched to dozens of GRBs within their $2\sigma$ error circle. Therefore, we cannot reliably claim that any of the GRB matches are real associations.

\subsubsection{Searching for Cataloged Optical Galaxy and Stellar Counterparts}\label{subsubsec:galaxysearch}

We searched for galaxy counterparts in the Panoramic Survey Telescope and Rapid Response System 1 \citep[PS1;][]{Chambers2016PS1} Source Types and Redshifts with Machine Learning catalog \citep[PS1-STRM;][]{Beck2021}, the Dark Energy Spectroscopic Instrument (DESI) Legacy Imaging Surveys DR8 photometric redshift catalog \citep[][]{Duncan2022}, SDSS DR18 \citep[][]{Almeida2023_SDSSDR18}, and the Galaxy List for the Advanced Detector Era + catalog \citep[][]{Dalya2022GLADEp}. We found nine transient candidates to be within $\sim$0.1-0.3\,arcsec of galaxies with photometric redshift estimates spanning $z_{\mathrm{phot}}\sim0.1-1.0$, shown in Table \ref{tab:RedshiftLnu}. We consider these to be real matches because of the small separations. In the PS1-STRM catalog, the matched counterparts (those with $z_{\mathrm{phot}}$) have high probabilities of being galaxies ($P\gtrsim0.89-0.99$) and not quasars, which is reassuring given that we have removed bright optically-selected AGNs. In the optical images, extended emission associated with galaxies can be seen (Figure \ref{fig:OpticalImages}) and the small separations from the radio positions suggest that the transient candidates could reside in the nuclei of the matched galaxies.

In Table \ref{tab:RedshiftLnu}, we also provide the derived (isotropic) spectral luminosity ($L_\nu$) and time between epoch 1 and 2 in the frame of the galaxy ($\Delta t/(1+z_{\mathrm{phot}})$). Note that we adopted $z_{\mathrm{phot}}$ with the smallest error when there are estimates from multiple catalogs. We did not propagate the redshift error to $L_\nu$ and we have not applied a k-correction (e.g., the factor of $(1+z)^{-(1+\alpha)}$ for a power law) due to our lack of knowledge on the exact spectral shape and the precise value of $\alpha$. Therefore, the derived $L_\nu$ should be treated as order-of-magnitude estimates over the \textit{observed} frequency range of 2.0--4.0\,GHz that could be slightly overestimated. The error of $\Delta t/(1+z_{\mathrm{phot}})$ is propagated from $z_{\mathrm{phot}}$. At the estimated redshifts, the spectral luminosities of our transient candidates are $L_{\nu}\sim10^{30}-10^{33}\,\mathrm{erg}\,\mathrm{s}^{-1}\,\mathrm{Hz}^{-1}$.

We note that five of our candidates are within the SDSS footprint, but only the counterpart of VT\,J0535+8345 was detected by SDSS, and with no spectrum. The counterpart of VT\,J0243-2840 has been observed by the 2dF Galaxy Redshift Survey \citep[2dFGRS;][]{Colless20012dFGRS} with a recorded spectroscopic redshift of $z=0.07$, which is the same value as the estimated photometric redshift. The 2dFGRS spectrum of VT\,J0243-2840 has been analyzed in detail by \citet{Somalwar2023J0243}, along with a newer optical spectrum. Thus, we do not perform any analysis on existing optical spectra of VT\,J0243-2840 and we simply reference the results of \citet[][]{Somalwar2023J0243} in relevant discussions. \citet{Kunert-Bajraszewska2025} also acquired optical spectra for three of our candidates -- VT\,J0243-2840, VT\,J0535+8345, and VT\,J2239-2231 -- but did not present the spectra, only referencing spectroscopic redshifts. For these three candidates, the quoted spectroscopic redshifts closely match the estimated photometric redshifts, and since their optical spectra do not appear to be public at the time of writing, we proceed with the photometric redshifts in our analyses.

We also searched for possible stellar counterparts in Gaia DR3 \citep[][]{GaiaMission2016,GaiaDR32023} but found no counterparts for candidates that are not matched to galaxies. Also, counterparts in the PS1-STRM catalogs have low probabilities of being stars. In addition, we searched the vicinity for stars with high proper motion that may have moved considerably between the optical and radio observations \citep[e.g.,][]{Driessen2023PM,Driessen2024SydneyStar}, but found no such cases. Therefore, our transient candidates do not appear to be associated with known or obvious optical stellar sources in the Milky Way. 

\begin{deluxetable*}{ccccc}
\tablecaption{Photometric Redshifts and and Derived Properties\label{tab:RedshiftLnu}}
\tablewidth{0pt}
\tablehead{
\colhead{Name} & \colhead{$z_{\mathrm{phot}}$} & \colhead{$L_{\nu,\mathrm{Epoch 1}}$} & \colhead{$L_{\nu,\mathrm{Epoch 2}}$} &
\colhead{$\Delta t/(1+z_{\mathrm{phot}})$} 
\\
\colhead{} & \colhead{} & \colhead{($10^{31}$\,erg\,s$^{-1}$\,Hz$^{-1}$)} & \colhead{($10^{31}$\,erg\,s$^{-1}$\,Hz$^{-1}$)} &
\colhead{(d)}
}
\startdata
VT\,J0013-3125 & 0.58$\pm$0.02 & 29.94  & 28.93 & 607.8$\pm$7.7  \\
VT\,J0243-2840 & 0.07$\pm$0.006 & 0.69   & 0.76  & 923.5$\pm$5.2  \\
VT\,J0435-2238 & 0.28$\pm$0.03 & 5.08   & 5.23  & 772.1$\pm$18.1 \\
VT\,J0535+8345 & 0.15$\pm$0.005 & 1.55   & 1.79  & 895.8$\pm$3.9  \\
VT\,J1400+6413 & 1.06$\pm$0.11 & 130.65 & 121.7 & 483.3$\pm$25.8 \\
VT\,J1757+2929 & 0.48$\pm$0.08 & 24.13  & 28.32 & 726.4$\pm$39.3 \\
VT\,J2210+4922 & 0.08$\pm$0.004 & 0.29   & 0.23  & 813.7$\pm$3.0  \\
VT\,J2239-2231 & 0.14$\pm$0.008 & 0.98   & 0.87  & 842.4$\pm$5.9  \\
VT\,J2309-1752 & 0.09$\pm$0.011 & 0.59   & 0.39  & 866.4$\pm$8.7 
\enddata
\tablecomments{Values after the $\pm$ symbol are $1\sigma$ errors.}
\end{deluxetable*}

\subsubsection{Searching for Additional Infrared, X-ray, and Optical
Counterparts}\label{subsubsec:opticalIRsearch}

We searched for IR counterparts in the Wide-field Infrared Survey Explorer \citep[WISE;][]{Wright2010WISE} catalogs -- AllWISE \citep[][]{Cutri2014AllWISE}, unWISE \citep[][]{Schlafly2019unWISE}, and CatWISE \citep[][]{Marocco2021catWISE}. WISE magnitudes of 11 for our transient candidates from cross-matching with AllWISE (using a match radius of 1.5\,arcsec) are shown in Table~\ref{tab:WISEMag}, which we use for later analyses regarding possible classifications (Section~\ref{subsec:possibleclass}). None of these AllWISE matches were flagged for significant variability ($var\_flg=0$ for 10 candidates and $var\_flg=3$ for one candidate). We note that for a few candidates (VT\,J1506-2144, VT\,J1757+2929, and VT\,J2210+4922), we found AllWISE matches within $\sim$1.5--3.0\,arcsec, but we do not use these for our analyses because of possible issues from crowding and blending of nearby emission as seen in the images. Additionally, three of our candidates (VT\,J0339+3837, VT\,J1835+0457, and VT\,J1906+0849) with no counterparts in AllWISE have relatively faint counterparts in unWISE or CatWISE.

We searched for X-ray counterparts in the Chandra Source Catalog Release 2.0 \citep[][]{Evans2010CSC}, the XMM-Newton Serendipitous Source Catalog DR13 \citep[][]{Webb20204XMM}, the Swift-XRT Point Source Catalog \citep[][]{Evans20202SXPS,Evans2023LSXPS} and the eROSITA-DE DR1 catalogs \citep[v1.1 for Main and v1.0 for Hard;][]{Merloni2024eROSITADR1}. We found an X-ray counterpart for VT\,J0232+3527 observed by Chandra (Obs ID 7111) with a reported X-ray flux of $F_{\mathrm{0.5-7.0keV}}\sim2\times10^{-14}\,\mathrm{erg}\,\mathrm{s}^{-1}\,\mathrm{cm}^{-2}$. We also found X-ray observations for VT\,J0243-2840 observed by Swift (obsid 00014300001 and 00014364003) and XMM-Newton (Obs. ID 0872393201) obtained by \citet{Somalwar2023J0243} specifically for this radio transient candidate. \citet{Somalwar2023J0243} reported an X-ray flux of $F_{\mathrm{0.3-10keV}}\sim10^{-13}\,\mathrm{erg}\,\mathrm{s}^{-1}\,\mathrm{cm}^{-2}$ (corresponding to $L_{\mathrm{0.3-10keV}}\sim10^{42}\,\mathrm{erg}\,\mathrm{s}^{-1}$) with a soft photon index of $\Gamma_{\mathrm{X}} \sim3$ and a negligible intrinsic column density. Note that based on the data archives, most of our other transient candidates have never been in any observing fields of Chandra, XMM-Newton, or Swift-XRT.

In eROSITA-DE DR1, only five of our transient candidates (VT\,J0243-2840, VT\,J0435-2238, VT\,J1455-2014, VT\,J1506-2144, VT\,J1659-3223) are within the covered hemisphere, and only VT\,J0243-2840 was detected at $F_{\mathrm{0.2-2.3keV}}\sim10^{-13}\,\mathrm{erg}\,\mathrm{s}^{-1}\,\mathrm{cm}^{-2}$ (confirming the soft spectrum). For those not detected in eROSITA-DE, we found $3\sigma$ upper limits\footnote{Accessed through \url{https://erosita.mpe.mpg.de/dr1/erodat/upperlimit/single/}. Details of the upper limits are described by \citet{Tubin-Arenas2024eROSTIADR1Limits}.} to be $F_{\mathrm{0.2-2.3keV}}\lesssim(4-9)\times10^{-14}\,\mathrm{erg}\,\mathrm{s}^{-1}\,\mathrm{cm}^{-2}$. For VT\,J0435-2238, which has a photometric redshift estimate, the upper limit translates to $L_{\mathrm{0.2-2.3keV}}\lesssim10^{42}\,\mathrm{erg}\,\mathrm{s}^{-1}$. Overall, we found fairly few X-ray constraints for our transient candidates. 

For transient candidates that are not matched to galaxy counterparts, we searched the PS1 survey \citep[][]{Chambers2016PS1} and the DESI Legacy Surveys DR10 \citep[][]{Dey2019DESILS} for unclassified optical counterparts (see Figure \ref{fig:OpticalImages} for cutout optical images). Note for one candidate at low declination (VT\,J1659-3223) that is not within the coverage of PS1 and Legacy Survey DR10, we instead searched Gaia DR3 and the SkyMapper Southern Sky Survey (SMSS) DR2 \citep[][]{Onken2019SMSSDR2}. Also note that we only consider counterparts separated by $<$1\,arcsec from the radio positions \citep[similar to the VLASS astrometric uncertainty;][]{Gordon2021VLASS}. From this search, we found three candidates (VT\,J0232+3527, VT\,J1506-2144, and J1906+0849) with faint and red optical counterparts, which leaves four candidates (VT\,J0404+4205, VT\,J1659-3223, VT\,J1825-2545, and VT\,J2251+5201) with no optical or IR counterparts.

Finally, we checked the Zwicky Transient Facility public DR19 \citep[][]{Bellm2019ZTF,Masci2019ZTF} and the Asteroid Terrestrial-impact Last Alert System forced photometry \citep[][]{Tonry2018ATLAS,Smith2020ATLASTransient,Shingles2021ATLASWeb} for optical light curves at the location of our transient candidates. We found no evidence of variability in the optical light curves for any candidate.

\subsection{General Radio Properties}\label{subsec:generalradioproperty}

\subsubsection{Spectral Energy Distributions} \label{subsubsec:radioSEDs}

In Figure \ref{fig:RadioSEDs}, we show crude radio SEDs of our inverted-spectra transient candidates using flux density measurements and upper limits from Table \ref{tab:candobstable}. We show a power law through the VLASS and VCSS measurements with a shaded error region for the one candidate detected in both surveys. For the other 20 cases with no VCSS detections, we show a shaded region that spans the range from a shallow power law with the lower limit of $\alpha$ (through the VCSS upper limit) to a steep power law with $\alpha=2.5$, the expected spectral index for SSA (in a homogeneous environment).

In our search for potential transients, non-detection in NVSS was considered a primary indicator of significant variability. As shown in the radio SEDs, connecting the VLASS detection with the NVSS non-detection would require $\alpha\gtrsim 2.5$, which would be highly unusual for persistent radio sources \citep[e.g., very few compact VLASS sources have $\alpha>2$ as seen by][]{Gordon2021VLASS}. The natural explanation would be variability between NVSS and VLASS, which is confirmed by the RACS-mid \textit{detections} all being brighter than the NVSS $3\sigma$ upper limits. However, there may still be a possibility that we have selected variable radio sources with highly inverted spectra, which we also discuss below.

The crude radio SEDs with measurements at different frequencies provide rough depictions of the spectral shape if we assume that our transient candidates evolve slowly at the observed frequencies (which is likely a reasonable assumption based on the two epochs of VLASS; see Section~\ref{subsubsec:radiovar}). Broadly, we notice two types of spectral shape over $\nu\sim 0.8-3.0\,\mathrm{GHz}$: (i) highly inverted spectra and (ii) peaked spectra. The highly inverted spectra are seen with an approximate slope of $\alpha \sim 2.0-3.0$ connecting the VLASS and RACS measurements (e.g., VT\,J0013-3125). On the other hand, the peaked spectra show RACS measurements comparable to the VLASS measurements and above the VCSS measurements/limits (e.g., VT\,J0243-2840), indicating a turnover/peak near the observed frequencies ($\nu_{\mathrm{peak}}\lesssim 1-3\,\mathrm{GHz}$). In the final column of Table~\ref{tab:candobstable}, we list the broad spectral shape for each transient candidate. Overall, these spectra are consistent with the expected synchrotron spectra of transients that are highly inverted at low frequency and turn over at high frequency \citep[][]{Sari1998}. In this context, the difference between the observed highly inverted spectra and peaked spectra is the location of the spectral peak, i.e., the former suggests a spectral peak beyond the observed frequencies ($\nu_{\mathrm{peak}}\gtrsim 3\,\mathrm{GHz}$). The exact location of the spectral peak depends on various physical parameters, which we discuss in Section~\ref{subsec:physicsinvertedspectrum}.

For a few of our transient candidates, we also notice two unusual characteristics or complexities in their radio SEDs. The first is the LoTSS detections of VT\,J0232+3527 and VT\,J1757+2929, which are brighter than the RACS-low measurements, possibly revealing excess emission at low frequency on top of the inverted spectrum (power law) traced by RACS. An explanation for such excess could be a different spectral component at low frequency (e.g., a power law with $\alpha < 0$ at $\lesssim$800\,MHz), perhaps associated with persistent radio emission from the host galaxy \citep[e.g.,][]{Callingham2017} or a separate low-frequency transient component. Due to limited spectral and temporal coverage, the exact characteristics (e.g., spectral shape) and origin of the low-frequency emission is currently unclear. Nonetheless, it is intriguing to note that since two out of four LoTSS measurements reveal a low-frequency spectrum more complex than a simple power law, we may speculate that such complexity could be common among our transient candidates.

\begin{figure*}
\epsscale{1.14}
\plotone{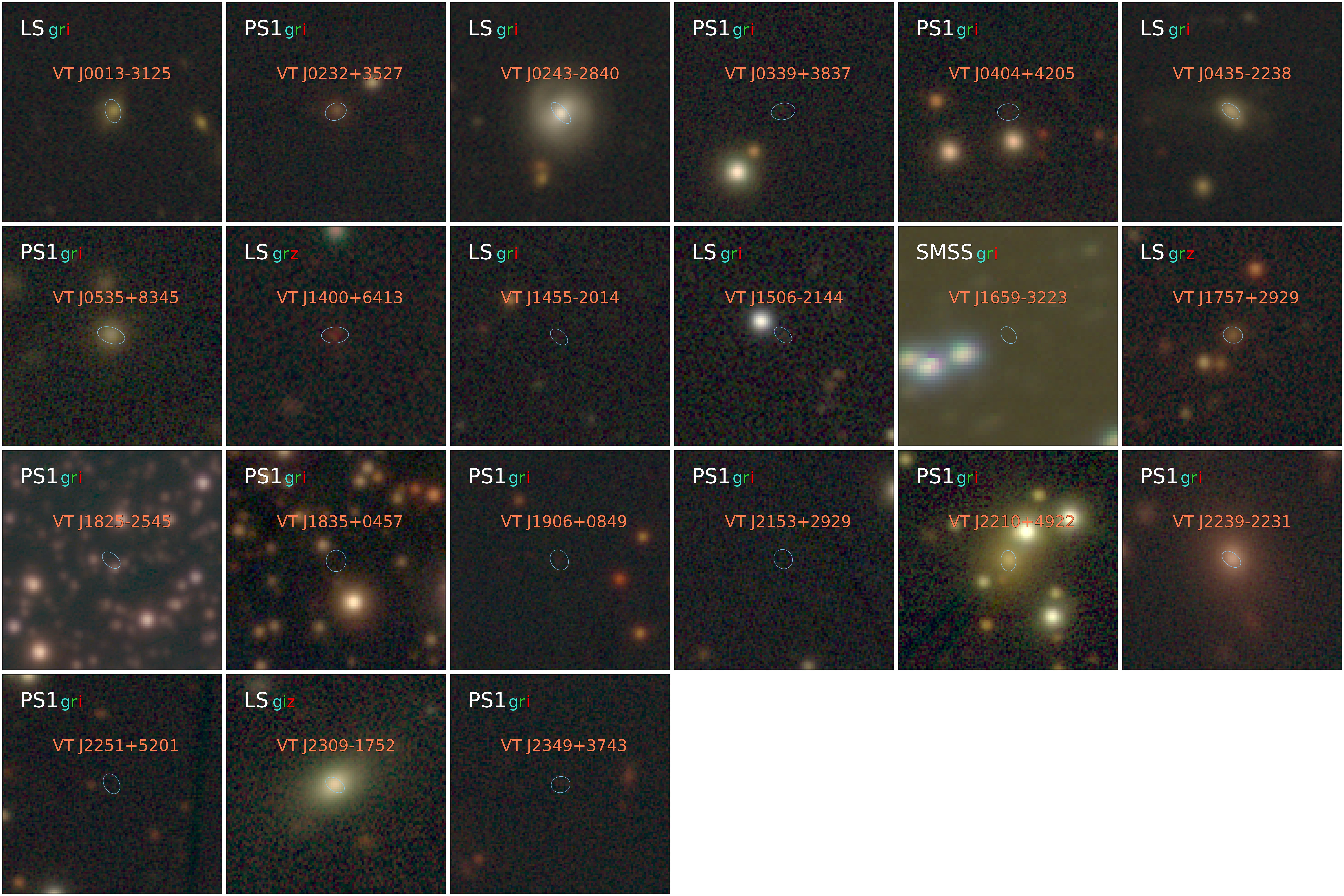}
\caption{Three-color optical cutout images (30$\times$30\,arcsec) of the inverted-spectra transient candidates from DESI Legacy Survey (LS) DR10, PS1, or SMSS DR2. A combination of the $griz$ bands, depending on availability, are used to for the three colors. The names and the synthesized radio beam (blue ellipse) of the candidates are shown in each image.\label{fig:OpticalImages}}
\end{figure*}

\begin{deluxetable*}{ccccc}
\tablecaption{AllWISE Magnitudes of Transient Candidates\label{tab:WISEMag}}
\tablewidth{0pt}
\tablehead{
\colhead{Name} & \colhead{W1 (mag)} & \colhead{W2 (mag)} & \colhead{W3 (mag)} &
\colhead{W4 (mag)} 
}
\startdata
VT\,J0013-3125 & 15.94$\pm$0.05 & 15.55$\pm$0.11 & $>$12.25         & $>$8.94         \\
VT\,J0232+3527 & 14.49$\pm$0.03 & 13.93$\pm$0.04 & 11.37$\pm$0.17 & $>$8.18         \\
VT\,J0243-2840 & 14.14$\pm$0.03 & 14.03$\pm$0.03 & 10.60$\pm$0.07 & 7.68$\pm$0.18 \\
VT\,J0435-2238 & 15.24$\pm$0.03 & 14.77$\pm$0.06 & 11.53$\pm$0.17 & $>$8.60         \\
VT\,J0535+8345 & 14.10$\pm$0.03 & 13.51$\pm$0.03 & 11.09$\pm$0.12 & 8.59$\pm$0.30 \\
VT\,J1400+6413 & 16.03$\pm$0.04 & 15.72$\pm$0.08 & $>$13.07         & $>$9.31         \\
VT\,J1455-2014 & 16.83$\pm$0.11 & 16.57$\pm$0.38 & 12.44$\pm$0.43 & $>$8.97         \\
VT\,J2153+2929 & 17.78$\pm$0.19 & $>$17.55         & $>$12.17         & $>$8.76         \\
VT\,J2239-2231 & 13.35$\pm$0.03 & 13.15$\pm$0.03 & 12.04$\pm$0.29 & $>$8.32         \\
VT\,J2309-1752 & 14.02$\pm$0.03 & 13.78$\pm$0.04 & 10.48$\pm$0.09 & 8.06$\pm$0.24 \\
VT\,J2349+3743 & 16.44$\pm$0.07 & 15.92$\pm$0.14 & $>$12.19         & $>$8.93       
\enddata
\tablecomments{Magnitudes are in the Vega system. Values after the $\pm$ symbol are $1\sigma$ errors. The $>$ sign indicates upper limit magnitude.}
\end{deluxetable*}

The other unusual characteristic is the extremely inverted ($\alpha\gtrsim4.0$) spectrum between RACS and VLASS seen in VT\,J1835+0457. In particular, the RACS-mid non-detection for VT\,J1835+0457 is fully consistent with the NVSS non-detection, meaning that we have no evidence of significant variability over the past few decades for this source. Therefore, rather than a radio transient with an optically thick $\alpha > 2.5$ spectrum \citep[which can happen, e.g., due to external FFA;][]{Sfaradi2023}, it is entirely possible that VT\,J1835+0457 is a persistent source, i.e., a rare case of a radio-loud AGN with an extremely inverted spectrum \citep[e.g.,][]{Callingham2015}. In addition to this source, we also note that a number of our transient candidates have RACS-mid flux densities that are only slightly brighter than the NVSS $3\sigma$ upper limits. These candidates could perhaps also be explained by radio-loud AGNs with highly inverted spectra ($\alpha > 2.5$) that were below the depth of NVSS but brightened moderately prior to RACS-mid at 1.4\,GHz. The existence of such sources certainly raises the concern that our sample of transient candidates may still be ``contaminated'' by variable AGNs that have unusually inverted spectra. We discuss this possibility in more detail in Section~\ref{subsubsec:AGNVar} and consider it as a caveat of our method in Section~\ref{subsec:feasibilitymethod}.

Finally, we note that all of our transients candidates have been detected in the two epochs of VLASS and many are detected in other radio surveys at different frequencies over several epochs. We consider these detections to be strong evidence that our transient candidates are real objects and not imaging artifacts. 

\subsubsection{Radio Variability in VLASS}\label{subsubsec:radiovar}

\begin{figure*}
\epsscale{1.07}
\plotone{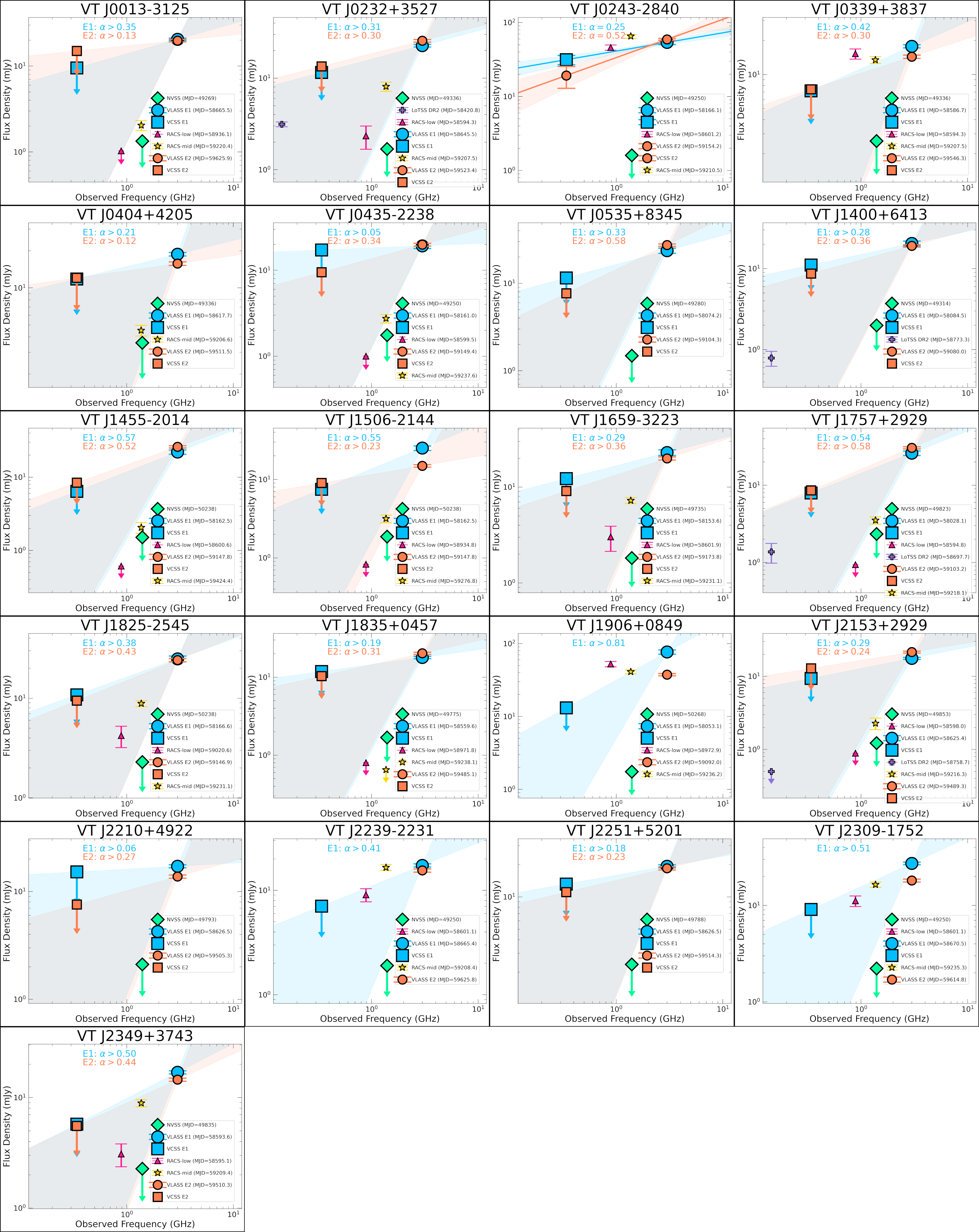}
\caption{Radio SEDs of the inverted-spectra transient candidates using measurements from NVSS, VLASS, VCSS, RACS-low, and LoTSS DR2 (see Table \ref{tab:candobstable}). Names of the candidates are shown as the plot titles. Spectral indices $\alpha$ from epoch 1 (E1) and epoch 2 (E2) are also shown at the top. Downward arrows below the points indicate ($3\sigma$) upper limits. Legend labels are ordered from top to bottom by increasing epoch. If the candidate is detected in both VLASS and VCSS, a power law with a shaded error region is shown. If the VCSS measurement is an upper limit, then the shaded region spans the range from a shallow power law with the lower limit $\alpha$ (through the upper limit of VCSS) to a steep power law with $\alpha=2.5$.\label{fig:RadioSEDs}}
\end{figure*}

\begin{figure}
\epsscale{1.14}
\plotone{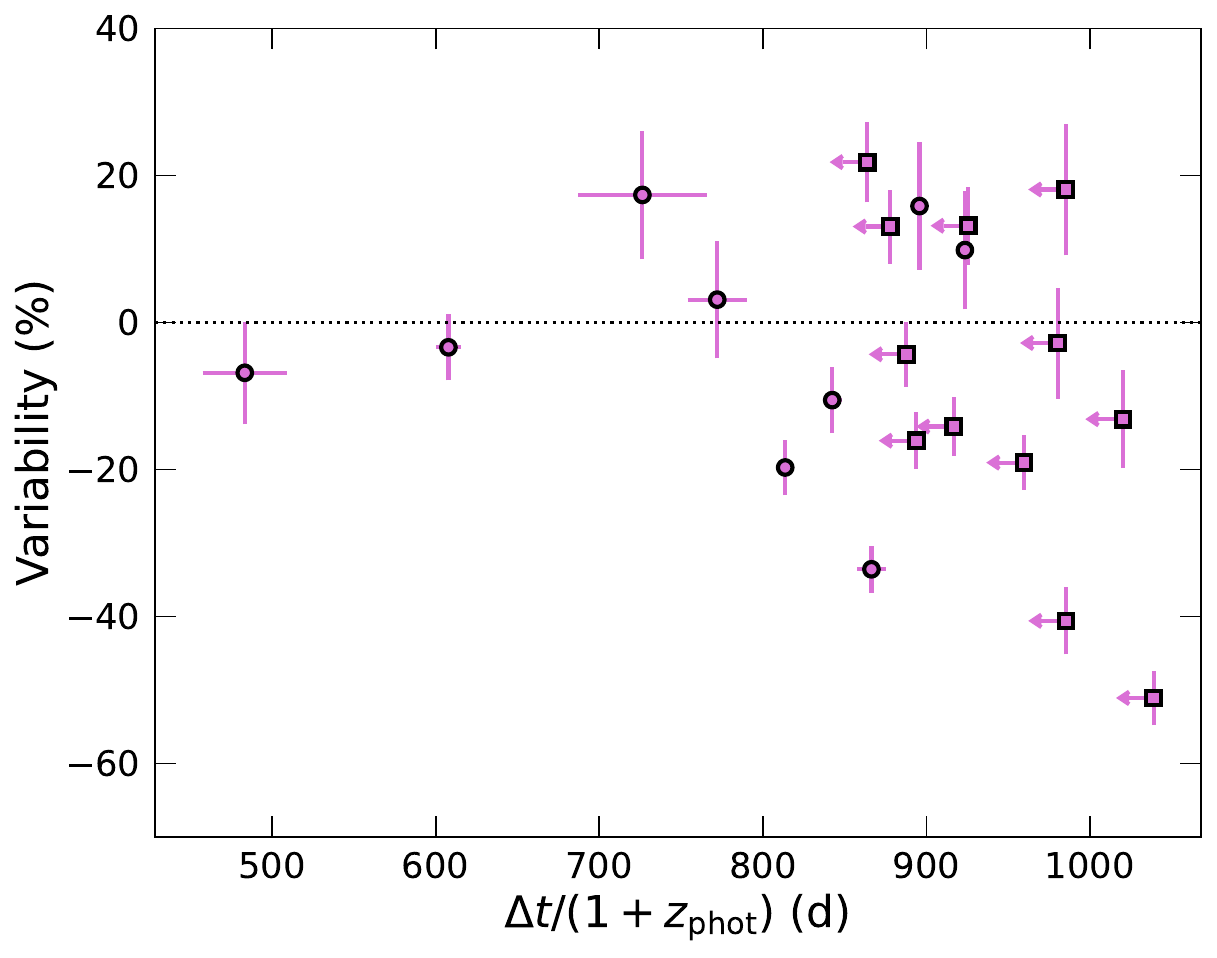}
\caption{Variability of the inverted-spectra transient candidates in VLASS, defined as the percentage change in $S_{3.0\mathrm{GHz}}$ between epoch 1 and 2, plotted against the change in time $\Delta t/(1+z_{\mathrm{phot}})$. Calculated $\Delta t/(1+z_{\mathrm{phot}})$ values are shown as circles and upper limits (due to the lack of redshift estimates) are shown as squares. \label{fig:Variability}}
\end{figure}

With flux density measurements from epoch 1 and 2 of VLASS, we calculated the variability of our inverted-spectra transient candidates, shown in Figure \ref{fig:Variability}. The majority of our candidates show $\lesssim$20\% variability in VLASS between epoch 1 and 2. Any brightening of the candidates were $\lesssim$20\% while the fading of some candidates reached $\sim$20--60\%. We note that only eight candidates show $>$3$\sigma$ variability, and that the variability of the other 13 candidates is not statistically significant. Overall, if these candidates are real transients, our results imply that we may be selecting sources that evolve very slowly over years to decades.

\begin{figure}
\epsscale{1.07}
\plotone{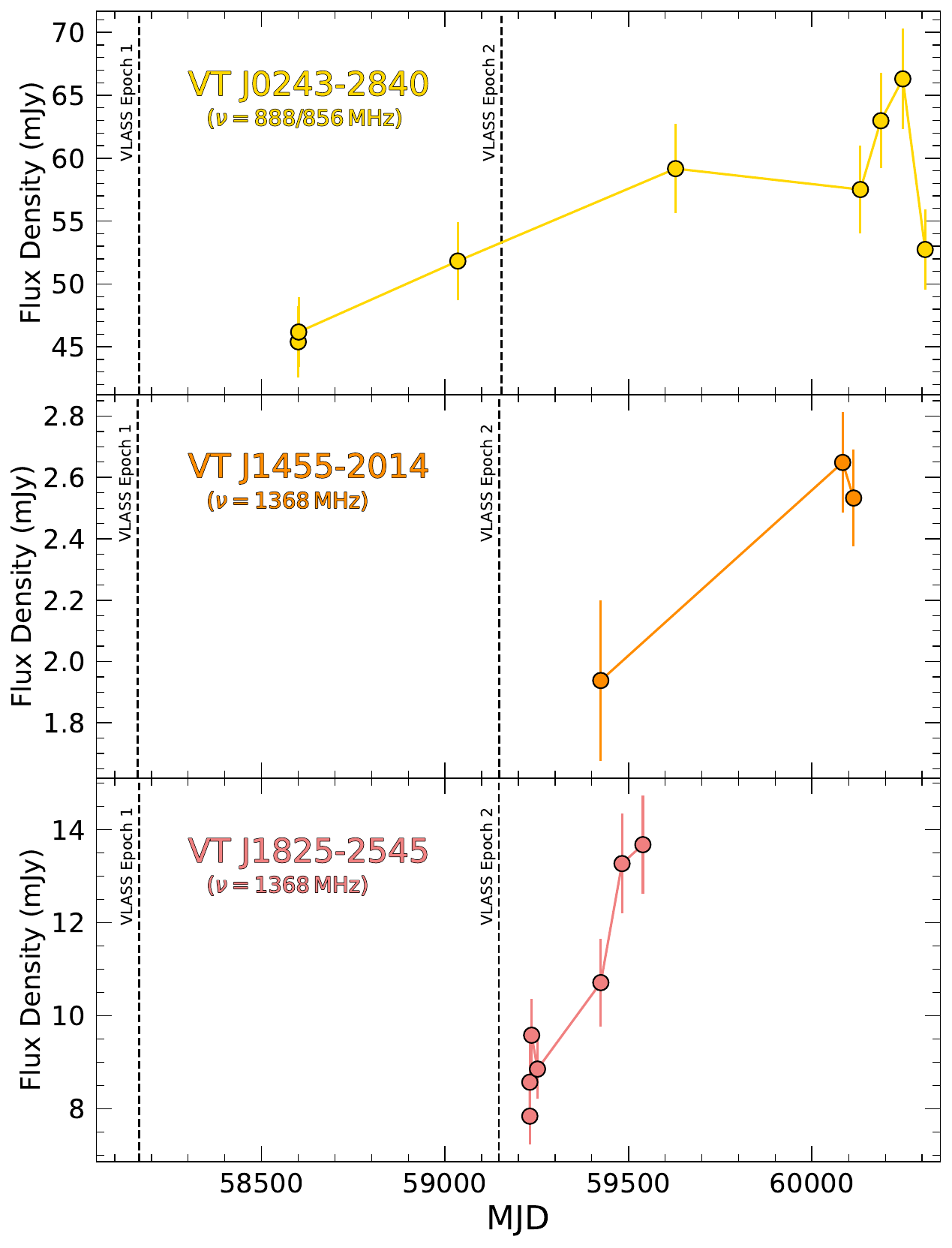}
\caption{Light curves of transient candidates with more than two data points at a similar frequency from ASKAP surveys RACS, VAST, FLASH, and WALLABY (see text for references). Frequency is shown below the Name. Dashed vertical lines mark the times of VLASS epoch 1 and 2. Note that the errors are conservative estimates that add in quadrature the cataloged flux density error, the image rms, and a systematic error of 6\%. \label{fig:ASKAPLC}}
\end{figure}

However, with two data points, it is ambiguous whether any variation between the two epochs occurred smoothly or sporadically. While changes in brightness are expected to follow some power law for a simple synchrotron transient \citep[][]{Granot&Sari2002}, irregular variation can also arise due to episodic flaring or extrinsic scattering \citep[e.g.,][]{Bannister2016}. Hence, the characteristic of the light curve is useful information that is reflective of the nature of the transient. For multi-epoch sampling, we checked ASKAP surveys (as of Mar. 15, 2024) and found light curves (with more than two data points) for VT\,J0243-2840, VT\,J1455-2014, and VT\,J1825-2545 from a combination of RACS, VAST, the First Large Absorption Survey in HI \citep[FLASH;][]{Allison2022FLASH}, and the Widefield ASKAP L-band Legacy All-sky Blind surveY \citep[WALLABY;][]{Koribalski2020WALLABY}. The light curves are shown in Figure~\ref{fig:ASKAPLC}. For these three transient candidates, there is some gradual brightening over years at a frequency lower than 3.0\,GHz (of VLASS), and we see no significant indication of any sporadic variation or multiple peaks beyond the measurement uncertainties. The only notable feature is the apparent narrow peak at the end of the VT\,J0243-2840 light curve, but it is hard to interpret due to large uncertainty. The overall smooth brightening is consistent with a synchrotron transient that predicts brightening at the inverted portion of the spectrum until the spectral peak moves below the observing frequency \citep[][]{Granot&Sari2002,Metzger2015}. At the moment, with no evidence of sporadic changes or multiple flares, slow evolution is still the simplest explanation for the marginal level (or even the lack) of variability of our transient candidates.

\section{Discussion} \label{sec:discussion}

We have selected a sample of 21 inverted-spectra transient candidates in epoch 1 of VLASS and VCSS and have compiled flux density measurements from recent radio surveys at different frequencies over similar epochs. Furthermore, we searched for counterparts in the optical, IR, and X-ray, and obtained photometric redshift estimates for a number of our candidates. In this section, we combine our findings and consider possible classifications of our transient candidates (Section~\ref{subsec:possibleclass}), examine physical constraints encoded within the inverted spectra (Section~\ref{subsec:physicsinvertedspectrum}), and reflect on the feasibility of our search method (Section~\ref{subsec:feasibilitymethod}).

\subsection{Considerations on Classifications} \label{subsec:possibleclass}

Based on the radio brightness/luminosity, timescale, spectral characteristics, as well as multi-wavelength associations, we can make some evidence-based speculations regarding the possible classifications of our transient candidates. For this exercise, the luminosity and timescale turn out to be two particularly useful pieces of information. Therefore, to begin, we show in Figure~~\ref{fig:SpecLumLC} the radio spectral luminosity light curves of our transient candidates (with a common arbitrary start time), and refer to it in relevant discussions below.

\subsubsection{Non-Relativistic Extragalactic Transients}

\begin{figure*}
\epsscale{1.14}
\plotone{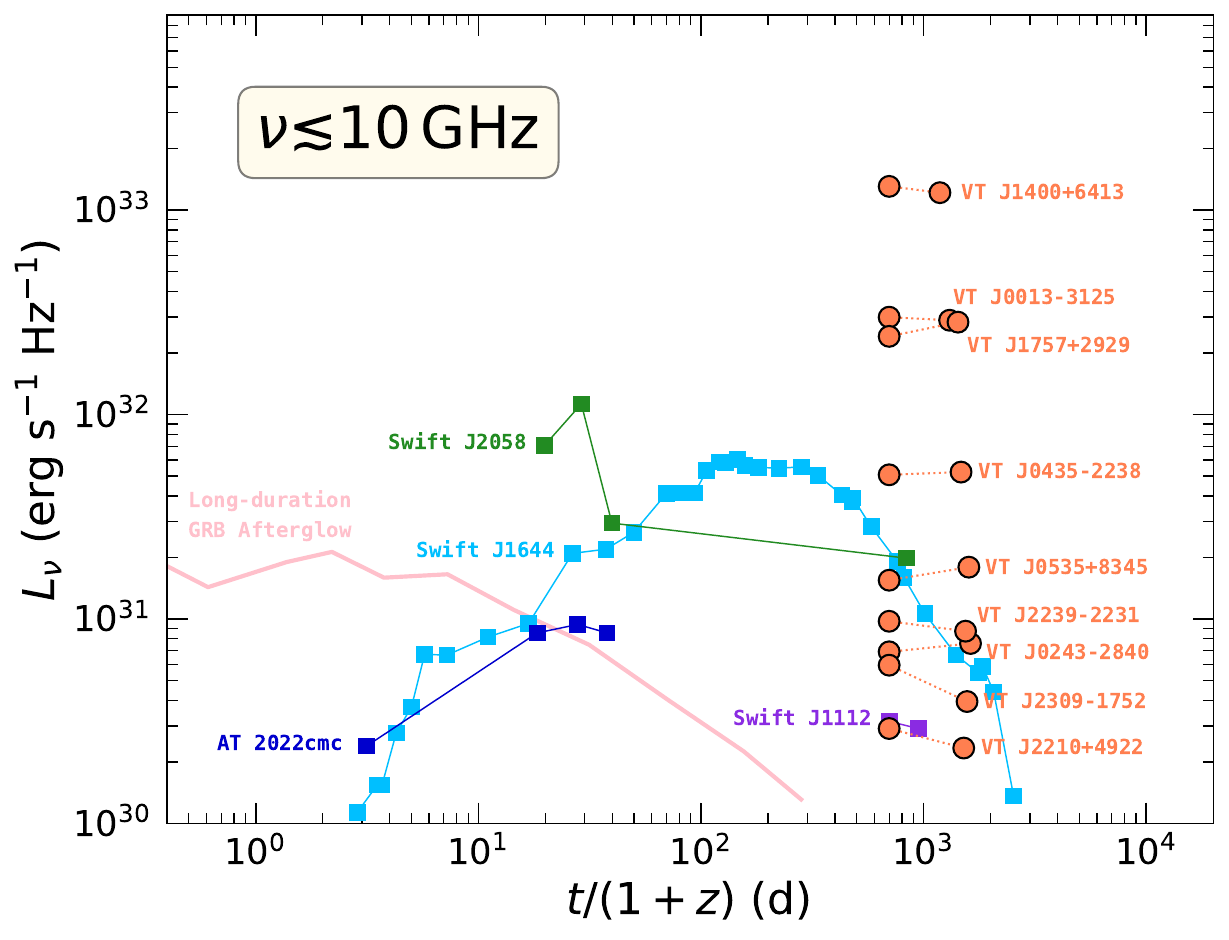}
\caption{Spectral luminosity light curves (in the frame of the host galaxies) from VLASS epoch 1 and 2 measurements (orange circles at 3.0\,GHz) of inverted-spectra transient candidates. Note that an \textit{arbitrary} start time (700 days prior to epoch 1) is set for all of our candidates since the real start times are not known. For comparison, we also show the averaged long-duration GRB afterglow light curve \citep[pink solid line at 8.5\,GHz;][]{Chandra2012} and light curves of four relativistic TDEs, Swift\,J1644 \citep[light blue squares at 5\,GHz except for last four points being 3\,GHz;][]{Berger2012Sw1644,Zauderer2013Sw1644,Eftekhari2018Sw1644,Cendes2021J1644}{}{}, Swift\,J2058 \citep[green squares at 7--9\,GHz;][]{Cenko2012SwJ2058,Pasham2015SwJ2058,Brown2017SwJ1112SwJ2058}, Swift\,J1112 \citep[purple squares at 5\,GHz;][]{Brown2017SwJ1112SwJ2058}, and AT\,2022cmc \citep[dark blue squares at 5\,GHz;][]{Andreoni2022AT2022cmc,Rhodes2023AT2022cmc}{}{} \label{fig:SpecLumLC}}
\end{figure*}

For our transient candidates with estimated photometric redshifts, the derived spectral luminosities span the range of $L_\nu \sim 10^{30}-10^{33}\,\mathrm{erg}\,\mathrm{s}^{-1}\,\mathrm{Hz}^{-1}$ (Figure~\ref{fig:SpecLumLC}). This level of brightness immediately disfavors supernovae (SNe) that have $L_\nu \lesssim 10^{29}\,\mathrm{erg}\,\mathrm{s}^{-1}\,\mathrm{Hz}^{-1}$ over $\nu\sim4-10\,\mathrm{GHz}$ \citep[][]{Bietenholz2021RadioSN} and luminous fast blue optical transients (LFBOTs) that have $L_\nu\lesssim 10^{30}\,\mathrm{erg}\,\mathrm{s}^{-1}\,\mathrm{Hz}^{-1}$ at $\nu\lesssim10\,\mathrm{GHz}$ \citep[][]{Ho2019,Margutti2019,Coppejans2020,Ho2020,Bright2022,Ho2022}. Also, SNe and LFBOTs typically evolve and fade significantly over $\sim$1000\,days at a few GHz, but our transient candidates only showed marginal levels of variability. Based on the brightness, we also disfavor thermal (non-relativistic) TDEs\footnote{We follow the conventional assumption that thermal TDEs produce much fainter radio emission than relativistic TDEs. However, we note that \citet{Somalwar2023TDEI} have recently reported evidence that there may not exist a clear boundary in radio luminosity between thermal and relativistic TDEs.} that have $L_\nu\lesssim 10^{30}\,\mathrm{erg}\,\mathrm{s}^{-1}\,\mathrm{Hz}^{-1}$ at $\nu\lesssim10\,\mathrm{GHz}$ \citep[e.g.,][]{Horesh2021ASASSN15oi,Cendes2023}. Thus, we do not consider non-relativistic extragalactic transients as plausible classifications.

\subsubsection{Gamma-Ray Bursts}

GRBs are extragalactic relativistic transients that can produce bright radio afterglows with $L_\nu > 10^{30}\,\mathrm{erg}\,\mathrm{s}^{-1}\,\mathrm{Hz}^{-1}$ \citep[e.g.,][]{Chandra2012}. Although we did not find any reliable GRB associations for our transient candidates (Section~\ref{subsec:archivetransient}), such an association is still possible given the large positional uncertainties of GRBs and potential cases of off-axis GRBs \citep[e.g.,][]{Law2018}. However, there are some inconsistencies between the observed properties of our transient candidates and the expected evolution of GRB afterglows.

For our transient candidates, the persistent spectral luminosity on the level of $L_\nu \sim 10^{30}-10^{33}\,\mathrm{erg}\,\mathrm{s}^{-1}\,\mathrm{Hz}^{-1}$ over $\sim$years to a decade timescale is unusual in the context of GRB afterglows, which typically evolve substantially over days to months \citep[][]{Pietka2015}. For example, the observed averaged long-duration GRB afterglow light curve maintains $L_\nu \sim 10^{31}\,\mathrm{erg}\,\mathrm{s}^{-1}\,\mathrm{Hz}^{-1}$ only over the first month, and fades by more than an order of magnitude over a few hundred days \citep[][also shown Figure~\ref{fig:SpecLumLC}]{Chandra2012}. Therefore, the light curves of typical long-duration GRB afterglows are generally inconsistent with our transient candidates even if the afterglows were detected in VLASS epoch 1 at an early stage. We note that short-duration GRB afterglows are fainter than long-duration GRB afterglows \citep[][]{Fong2015}, so the same inconsistencies exist. 

Also, the radio SEDs of our transient candidates show $\nu_{\mathrm{peak}}\gtrsim\mathrm{GHz}$ for $\gtrsim$3\,yr, which is difficult to explain with GRB afterglows that usually peak at much lower frequencies at such late times. As an example, we estimate $\nu_{\mathrm{peak}}$ using the standard GRB afterglow model by \citet{Granot&Sari2002} that assumes a spherical adiabatic blast wave, and adopt reasonable parameters \citep[e.g.,][]{Bloom2003,Santana2014,Metzger2015,Beniamini2017,Chrimes2022} of isotropic kinetic energy $E_{\mathrm{K,iso}} = 10^{53}\,\mathrm{erg}$, fraction of energy in the electrons $\epsilon_e = 10^{-1}$, fraction of energy in the magnetic field $\epsilon_B = 10^{-4}$, electron power-law index $p=2.5$, and number density $n=1\,\mathrm{cm}^{-3}$ for a constant density interstellar medium. Taking $t_{\mathrm{obs}}=1000\,\mathrm{d}$ (observed time approximately between VLASS epochs) and $z=0.2$, we find $\nu_{\mathrm{peak}}\approx222\,\mathrm{MHz}$ at the self-absorption frequency. Therefore, the afterglow model predicts that radio emission probed by VLASS and VCSS should be optically thin with $\alpha < 0$ after $\sim$3\,yr, which is inconsistent with the observed inverted spectra of our transient candidates during epoch 2.

In similar fashion, we find discrepancies between our transient candidates and the theoretical predictions of GRB afterglows in VLASS by \citet{Metzger2015}. In their simulation, on-axis and off-axis long-duration GRB afterglows in VLASS are virtually all expected to brighten or fade by more than a factor of 2 between two VLASS epochs and only be detectable in a single epoch. Compared to the observed marginal levels of variability between VLASS epochs (Figure~\ref{fig:Variability}), this again confirms the inconsistency in the timescale between our candidates and GRB afterglows.

Lastly, we note that for the nine transient candidates matched to galaxies, the offsets between the radio positions and the optical galaxy centers are $\sim$0.1-0.3\,arcsec. Since the majority of the matched galaxies are at $z_{\mathrm{phot}}\lesssim0.2$, we find a preference for projected offsets of $r_{\mathrm{proj}}\lesssim0.2-0.8\,\mathrm{kpc}$. On the other hand, for long-duration (short-duration) GRBs, roughly $\lesssim$30\% ($\lesssim$10\%) of explosion sites are observed to be within $r_{\mathrm{proj}}\lesssim0.8\,\mathrm{kpc}$ \citep[e.g.,][]{Lyman2017,Fong2022}. This could be another discrepancy between our transient candidates and GRB afterglows, although we note that this discrepancy is only tentative due to uncertainties in the radio positions and $z_{\mathrm{phot}}$.

In summary, based on the observed persistent brightness and inverted spectra over $\gtrsim$3\,yr, we do not consider GRB afterglows as a favorable explanation for our transient candidates. However, we do not fully rule out GRB afterglows, because fine-tuning of explosion parameters \citep[perhaps highly energetic cases similar to those reported by][]{Cenko2011} and considerations of energy injection \citep[][]{Zhang2001,Zhang2002,Gao2013} and late-time non-relativistic evolution \citep[][]{Wygoda2011,Sironi2013} may lead to afterglow properties similar to our transient candidates. Detailed examination of models and viable parameter space is beyond the scope of this study, especially considering our lack of observational constraints.

\subsubsection{Galactic Transients}

A number of our transient candidates (not matched to galaxies) lie near the Galactic plane: VT\,J0404+4205, VT\,J1659-3223, VT\,J1825-2545, VT\,J1835+0457, VT\,J1906+0849, and VT\,J2251+5201, with Galactic latitudes of $|b|<8\,\mathrm{deg}$. VT\,J1906+0849 in particular has $b=0.74\,\mathrm{deg}$, perhaps suggesting a Galactic origin. Although extended sources (e.g., resolved Galactic sources) are excluded from our transient search, our selection criteria do not explicitly reject compact Galactic radio transients. Therefore, it is worth considering the possibility that some of our transient candidates are Galactic transients. 

Firstly, the radio brightness and lack of bright stellar associations observed for our transient candidates place heavy constraints on any Galactic stellar transient interpretation. For example, if radio flares from K- or M-type dwarfs can have $L_{\nu}\sim10^{15}\,\mathrm{erg}\,\mathrm{s}^{-1}\,\mathrm{Hz}^{-1}$ \citep[at $\sim$GHz;][]{Smith2005,Mooley2016CNSS}, then the flux densities of our transient candidates of $F_{\nu} \gtrsim15\,\mathrm{mJy}$ would place the stars at $d \lesssim 7.5\,\mathrm{pc}$. Since the coolest M-type dwarfs are found to have an average $r$-band absolute magnitude of $M_{r}\sim18.5$ \citep[][]{Best2018,Cifuentes2020}, the lack of bright optical counterparts in PS1 with apparent magnitudes $m_{r}\lesssim 18.5$ is evidence against radio flares from K- or M-type dwarfs. For brighter radio transients associated with, e.g., cataclysmic variable stars, Algol systems, and RS Canum Venaticorum systems that have $L_{\nu}\lesssim10^{20}\,\mathrm{erg}\,\mathrm{s}^{-1}\,\mathrm{Hz}^{-1}$ \citep[][]{Pietka2015,Mooley2016CNSS}, $F_{\nu} \gtrsim15\,\mathrm{mJy}$ would require $d\lesssim2.4\,\mathrm{kpc}$. Out to a distance of 2.4\,kpc, the PS1 survey (with limiting $r$-band magnitude of 23.2) should detect stellar objects down to $M_{r}\lesssim11.3$, which include all stars hotter and brighter than M-type dwarfs \citep[][]{Cifuentes2020}, although some non-detections may be explainable by significant dust extinction at low Galactic latitudes. Overall, the lack of bright optical stellar associations for our transient candidates disfavors many types of Galactic radio transients originating from stars.

In addition, inconsistencies may also exist in the timescale. Most ``slow'' stellar radio transients, including stellar flares, X-ray binaries, dwarf novae, and magnetar giant flares, evolve over a typical timescale of hours to weeks \citep[e.g.,][and references therein]{Gudel2002,Pietka2015,Mooley2016CNSS}. On the other hand, our transient candidates show little variability over $\gtrsim$3\,yr in VLASS. This inconsistency can be resolved if we assume that the two epochs of VLASS coincidentally captured two separate flares with similar brightnesses. However, such coincidences should not be common, and at least for one candidate near the Galactic plane (VT\,J1825-2545), a smooth brightening is observed over hundreds of days after VLASS epoch 2 (Figure~\ref{fig:ASKAPLC}). Therefore, the observed timescale also challenges various Galactic transient interpretations.

If any of our transient candidates are Galactic, they are more likely to be the brightest Galactic radio transients detected at larger distances such that any optical counterpart would be faint or obscured. Bright Galactic radio transients that can produce $L_{\nu}\gtrsim10^{20}-10^{22}\,\mathrm{erg}\,\mathrm{s}^{-1}\,\mathrm{Hz}^{-1}$ include some classical novae \citep[][]{Chomiuk2021}, X-ray binaries \citep[][]{Merloni2003,Tetarenko2016}, and magnetar giant flares \citep[][]{Gaensler2005,Cameron2005}. In particular, some classical novae produce synchrotron emission that can evolve over a timescale of $\gtrsim$1000\,d \citep[][]{Chomiuk2021}, which may explain the observed properties of our transient candidates. Therefore, we cannot rule out rare cases of bright and slow evolving Galactic radio transients for some of our transient candidates not matched to galaxies.

\subsubsection{Variable Active Galactic Nuclei} \label{subsubsec:AGNVar}

The radio properties of our transient candidates are reminiscent of peaked-spectrum (PS) sources such as gigahertz-peaked spectrum (GPS) sources and high-frequency peakers (HFPs), which are radio-loud AGNs characterized by concave radio spectra peaking at $\nu_{\mathrm{peak}}\sim0.5-5\,\mathrm{GHz}$ and $\nu_{\mathrm{peak}}\gtrsim5\,\mathrm{GHz}$, respectively \citep[][]{ODea1998,ODea2021}. PS sources have radio luminosities spanning $L_{\nu}\sim10^{31}-10^{36}\,\mathrm{erg}\,\mathrm{s}^{-1}\,\mathrm{Hz}^{-1}$ and compact radio linear sizes less than a few kpc, down to a few pc for HFPs \citep[e.g.,][]{An&Baan2012,Patil2020}. From the high-frequency peaks and inverted spectra, PS sources have been interpreted as dynamically young precursors of large-scale Fanaroff-Riley classes of galaxies \citep[][]{An&Baan2012}, jets confined by dense environments \citep[the frustration scenario;][]{vanBreugel1984,Wagner2011}, or ``short-lived'' episodic events \citep[e.g., activity lasting $\lesssim10^4\,\mathrm{yr}$ due to radiation pressure instability in the accretion disk;][]{Czerny2009}. The properties of PS sources are broadly similar to those of our transient candidates, and since many of our candidates are matched to the optical centers of galaxies (to within the radio positional uncertainty), a nuclear origin for the radio emission is plausible. Thus, we consider the possibility that we may have selected variable AGNs (or even persistent AGN in the case of VT\,J1835+0457) instead of transients in the traditional sense (e.g., explosive transients).

\begin{figure}
\epsscale{1.07}
\plotone{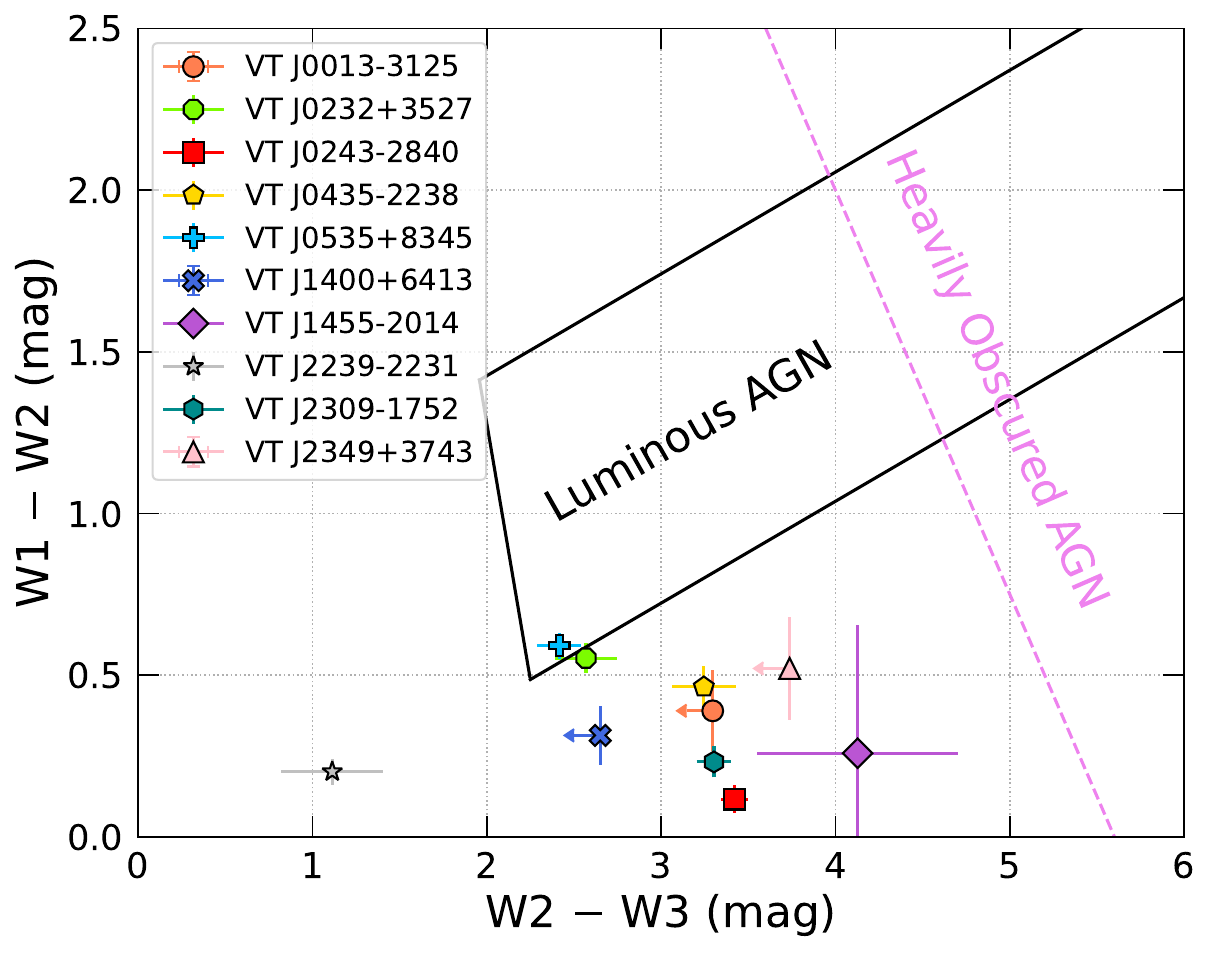}
\caption{AllWISE color-color diagram for our transient candidates. Upper limits of the W2--W3 color are marked with arrows. The labeled region for luminous AGNs is from \citet{Mateos2012}, and the region for heavily obscured AGNs is from \citet{Lonsdale2015} and \citet{Patil2020}. \label{fig:WISEColorColor}}
\end{figure}

Before discussing possible scenarios of AGNs, we first gather some information about classification based on the optical and IR counterparts of our transient candidates. In Figure~\ref{fig:WISEColorColor}, we show the AllWISE IR color-color diagram constructed using magnitudes from Table~\ref{tab:WISEMag} as well as regions containing luminous or heavily obscured AGNs. IR counterparts of our transient candidates generally fall outside of the AGN regions, which is expected since we have already removed cataloged AGNs and quasars during our transient search (Section~\ref{subsec:searchmethod}). The PS1-STRM classifications also indicate that the optical counterparts with photometric redshifts have low probabilities of being quasars. Together with the lack of significant optical and IR variability (Section~\ref{subsec:searchmultiwavecounterpart}), these findings confirm the general exclusion of AGNs with luminous optical and IR emission in our sample. We note two possible exceptions -- counterparts of VT\,J0232+3527 and VT\,J0535+8345 -- that are at the edge of the luminous AGN region. In particular, for VT\,J0232+3527, PS1-STRM does not report a photometric redshift but does report a high probability for quasar classification. 

Ideally, having X-ray information would be beneficial for classification, since X-ray emission is almost ubiquitous in AGNs \citep[][]{Padovani2017}. Unfortunately, we have very limited X-ray constraints for our transient candidates (Section~\ref{subsubsec:opticalIRsearch}), and existing upper limits do not rule out $L_{\mathrm{X}}\gtrsim10^{41}-10^{42}\,\mathrm{erg}\,\mathrm{s}^{-1}$ seen in AGNs \citep[][]{Ueda2014}. With the available information, we still cannot rule out weaker AGNs with fainter optical and IR emission and that were previously radio-quiet. For those lacking multi-wavelength information, we also cannot rule out distant AGNs (e.g., higher redshift quasars). In the following discussions, we consider if these types of AGNs can explain our transient candidates.

As noted in Section~\ref{subsubsec:radioSEDs}, our transient candidates show two broad types of spectral shape over $\sim$0.3--3.0\,GHz -- (I) highly inverted spectra and (II) peaked spectra. For the most inverted spectra, the observed differences between NVSS (upper limit) and RACS-mid at 1.4\,GHz are small, while for peaked spectra, significant brightening is seen. Therefore, the required types of variability are different, so we discuss the two cases separately below.

\medskip \noindent{\underline{(I) Highly Inverted Spectra} -- Moderate Variability}
\medskip

A significant fraction of our transient candidates (7 out of 22) show highly inverted spectra ($\alpha\gtrsim2.5$) with RACS-mid measurements that are only slightly brighter than the NVSS $3\sigma$ upper limits at 1.4\,GHz (Figure~\ref{fig:RadioSEDs}). For these candidates, a possible scenario is that a PS source slightly below the NVSS upper limit experienced moderate brightening of $\gtrsim$20\%--80\% prior to VLASS. This scenario may better explain the observed persistent brightness and inverted spectrum over $\gtrsim$3\,yr. There is also a single extreme case of VT\,J1835+0457 (with $\alpha\gtrsim4.0$) that shows a RACS-mid upper limit below the NVSS upper limit, meaning that it could be an extremely inverted persistent PS source. We note that the general exclusion of luminous AGNs and quasars does not disfavor PS sources, since many PS sources are classified as galaxies (not quasars) in the optical and IR \citep[][]{Hancock2009,Kunert-Bajraszewska2010,Kosmaczewski2020}.

For PS sources, variability is generally low at $\sim$GHz, at about a few percent over years \citep[][]{ODea1998,Fassnacht2001}, but a small fraction of PS sources vary by $\gtrsim$10\%--100\% \citep[e.g.,][]{Jauncey2003,Wu2013,Ross2021,Ross2022}. On $\sim$decade timescales, a large fraction of PS sources may show moderate levels of variability \citep[][]{Orienti2020}. We note that some variable PS sources are known to be misclassified flat-spectrum sources, likely blazars, that show peaked spectra during flares \citep[][]{Tinti2005,Torniainen2005,Orienti2007,Orienti2020}. However, a flat spectrum below NVSS would imply more than about an order of magnitude brightening at 3\,GHz in VLASS, which is much more drastic than typical blazar variability. Therefore, we disfavor blazar variability for our transient candidates. 

Intrinsic variability in PS sources may be related to adiabatic expansion or varying optical depth \citep[][]{Orienti2010,Ross2022,Orienti2020}. At 1.4\,GHz, the extrinsic effect of refractive interstellar scintillation (strong scattering) can also produce modulation of $m\lesssim60\%$ over days to months \citep[assuming a critical frequency $\nu_0\gtrsim4\,\mathrm{GHz}$ over a range of Galactic latitude;][]{Walker1998}. Another extrinsic effect is an extreme scattering event (ESE), which can lead to irregular variation in the flux density and spectral shape over weeks to months \citep[][]{Bannister2016}. These extrinsic effects may be a possible explanation, but the timing must be correct to produce a non-detection in NVSS and a detection in RACS-mid. Overall, regardless of the exact origin, moderate variability can explain many of our transients candidates with highly inverted radio spectra \textit{if} we make the assumption that their flux densities were \textit{just below} NVSS and that brightening was captured decades later. 

We also note that $\alpha\gtrsim2.5$ is quite unique for PS sources. Most PS sources have $\alpha \lesssim1.3$ in the optically thick segment \citep[][]{deVries1997,ODea1998,Snellen1998,Stanghellini1998,Orienti2012,Randall2012,Callingham2017}, and very few have $\alpha \gtrsim 2.0$. Although steep spectra at low frequencies are expected for homogeneous absorbers, inhomogeneity can result in flatter spectra for both SSA and FFA \citep[e.g.,][]{Tingay2003}, which is commonly seen in the radio structures of PS sources \citep[][]{ODea1998}. In extremely rare occasions, inverted spectra with $\alpha\gtrsim2.5$ have been found for PS sources, which are generally better described by inhomogeneous FFA rather than SSA \citep[][]{Callingham2015,Callingham2017,Keim2019,Mhaskey2019a,Mhaskey2019b}. Interpretations of these PS sources likely have significant implications on the ambient medium, possibly favoring the frustration scenario for some fraction of radio-loud AGNs \citep[][]{Mukherjee2016,Bicknell2018}. Also, irregular variability of ESEs can result in highly inverted spectra \citep[][]{Lazio2001}, which we cannot rule out with our limited temporal coverage. We speculate that the extreme case of VT\,J1835+0457 (with $\alpha\gtrsim4.0$) may be explained by significant FFA or an ESE if it is a radio-loud AGN. 

In the context of PS sources, the apparent overabundance of transient candidates with $\alpha\gtrsim2.0-3.0$ may be a selection effect related to our selection criteria of (i) non-detection in NVSS with $S_{\mathrm{1.4\mathrm{GHz}}}\lesssim 2.5\,\mathrm{mJy}$ and (ii) inverted radio spectra with $S_{\mathrm{3.0\mathrm{GHz}}}\gtrsim 15\,\mathrm{mJy}$. Assuming moderate flux density variability over decades but persistent spectral shape, these two criteria will only result in PS sources with highly inverted spectra.

In summary, many of our transient candidates with highly inverted spectra can be explained by PS sources that show moderate levels of variability (intrinsic or extrinsic) that just happened to produce non-detections in NVSS and brighten during the RACS-mid epoch. However, the highly inverted spectra imply that these PS sources must also be extremely rare, if our transient candidates are even AGNs in the first place. Considering the rarity of such cases, it is unclear at this point whether PS sources are a more favorable explanation over actual transients. If our transient candidates with highly inverted spectra are variable PS sources, we likely have selected a very peculiar and interesting population of AGNs, although they are persistent sources and not the intended target of this study. Follow-up observations are necessary for distinguishing between AGNs and transients and for constraining the physical origin of the variability, which we leave for future studies.

\medskip \noindent{\underline{(II) Peaked spectra} -- Extreme Brightening}
\medskip

A considerable number of our transient candidates (9 out of 22) show significant brightening of $\gtrsim$300\%--4000\% in RACS-mid at 1.4\,GHz compared to NVSS. They have peaked spectra near 1.4--3.0\,GHz, but some still show highly inverted spectra at $\lesssim$0.8-1.4\,GHz. This observed level of brightening cannot be explained by normal variability of PS sources. In the context of AGNs, these transient candidates are only consistent with recently discovered extreme cases of AGNs that have transitioned from radio-quiet to radio-loud over the past two decades in CNSS \citep[][]{Mooley2016CNSS,KunertBajraszewska2020,Wolowska2021} and VLASS \citep[][]{Nyland2020,Kunert-Bajraszewska2025}. The extreme brightening of AGNs in the radio has generally been associated with the launching of young jets possibly from episodic AGN activities, which may signify the birth or rejuvenation of PS sources.

Similarities in radio properties can be found between our transient candidates and the AGNs that experienced extreme brightening. These AGNs were undetected in FIRST or NVSS and brightened by more than a factor of few, up to more than an order of magnitude in CNSS or VLASS over $\sim$decade. The radio spectra of the brightened AGNs were characterized by peaked spectra with $\nu_{\mathrm{peak}}\gtrsim\mathrm{GHz}$ that displayed marginal variability over months and years after detection. In general, the radio properties of our transient candidates are very consistent with those of the extreme AGNs.

However, there are some potential uncertainties in associating our transient candidates with extreme AGNs. A significant number of extreme AGNs show $\alpha\lesssim1.0-2.0$ in the optically thick segment (perhaps for the same reason as typcial PS sources) while our transient candidates generally show $\alpha \gtrsim2.0$ between $\lesssim$0.8--1.4\,GHz, although there is a lack of constraints on $\alpha$ for those with $\nu_{\mathrm{peak}}\lesssim1-3\,\mathrm{GHz}$. In terms of galaxy classifications, the VLASS sample of extreme AGNs discovered by \citet{Nyland2020} was selected specifically from cataloged luminous AGNs and quasars, while the CNSS and VLASS samples discovered by \citet{Wolowska2021} and \citet{Kunert-Bajraszewska2025} contains both quasars and galaxies. On the other hand, we have explicitly removed luminous AGNs and quasars, and thus galaxies that do not show strong AGN signatures in the optical and IR should constitute a major fraction of our sample of transient candidates. 

For the galaxies, \citet{Wolowska2021} assumed that the radio brightening was due to AGN activity and did not consider alternative scenarios such as TDEs. A possible counterexample to this assumption is VT\,J0243-2840 in our sample, which was examined in detail by \citet{Somalwar2023J0243}. The host galaxy of VT\,J0243-2840 is a weak Seyfert galaxy, and based on the lack of strong AGN activity, low black hole mass, soft X-ray spectrum, and small radio linear size, \citet{Somalwar2023J0243} concluded that VT\,J0243-2840 would be an extremely unusual case if it was caused by AGN activity (although this was not ruled out). On the other hand, \citet{Somalwar2023J0243} found that a relativistic TDE is a plausible (perhaps more preferable) scenario for VT\,J0243-2840. However, we also note that \citet{Kunert-Bajraszewska2025}  instead favor the AGN interpretation for VT\,J0243-2840, as well as for VT\,J0535+8345 and VT\,J2239-2231. They argue in favor of weaker and more compact AGN jets that evolve in parallel with the known population of GPS and HFPs. Overall, we speculate that similar ambiguities may exist for our other transient candidates, so additional constraints will be necessary for establishing the origin of the radio brightening in galaxies.

In summary, based on the radio properties, AGNs transitioning from radio-quiet to radio-loud is a plausible explanation for our transient candidates that show significant brightening at 1.4\,GHz over $\sim$decade timescale. However, a major uncertainty lies in the classification of the host galaxy and whether or not there is ongoing AGN activity that can trigger such brightening. Lastly, we remark that phenomenologically, such extreme brightening of AGNs will appear as radio transient objects and thus will be identified in any transient search over $\sim$decade timescale unless excluded through extensive multi-wavelength constraints. While this may be non-ideal for searches targeting TDEs or explosive transients (which was the original intent of this study), transient AGNs are nonetheless interesting objects that may have important implications on topics including the evolution of radio-loud AGNs, jet launching from supermassive black holes, and the nature of changing-look AGNs \citep[][]{Wolowska2017,Nyland2020,Wolowska2021,Somalwar2023J0243,Meyer2025,Laha2025}. 

\subsubsection{Relativistic Tidal Disruption Events}

Jet launching has been observed in a small subset of TDEs known as relativistic TDEs or jetted TDEs, which generate bright long-lasting radio synchrotron emission as a result of a shock wave propagating through the circumnuclear medium \citep[][]{Alexander2020}. Currently, only a handful of on-axis relativistic TDEs has been discovered in the X-ray and optical, including Swift\,J1644+57 \citep[][]{Bloom2011,Burrows2011,Zauderer2011}{}{}, Swift\,J2058.4+0516 \citep[][]{Cenko2012SwJ2058}, Swift\,J1112.2-8238 \citep[][]{Brown2015}, and AT\,2022cmc \citep[][]{Andreoni2022AT2022cmc}. Radio light curves representative of these relativistic TDEs are plotted in Figure~\ref{fig:SpecLumLC}, showing $L_{\nu}\sim10^{30}-10^{32}\,\mathrm{erg}\,\mathrm{s}^{-1}\,\mathrm{Hz}^{-1}$ and timescales of at least several years. For the most well-studied case, Swift\,J1644+57, the radio spectral peak was above $\gtrsim$GHz for the observed period of about 2800 days \citep[or 2060 days in the frame of the host galaxy;][]{Cendes2021J1644}. Therefore, the radio properties of our transient candidates are consistent with observed relativistic TDEs. 

We note that on-axis relativistic TDEs are extremely rare, with a volumetric rate of $\mathcal{R}\approx 0.02\,\mathrm{Gpc}^{-3}\,\mathrm{yr}^{-1}$ inferred from the ZTF survey \citep[][]{Andreoni2022AT2022cmc} that is only $\sim$1\% of the thermal TDE rate even after a beaming correction \citep[][]{Teboul2023,Yao2023ZTFTDE}. Based on this rate, the estimated number of on-axis and off-axis relativistic TDEs in VLASS would be
\begin{eqnarray}
    N_{\mathrm{VLASS}} &\approx& \frac{\mathcal{R} \Delta t \Delta \Omega C }{f_{b}}\frac{4\pi d^3}{3} \nonumber\\ &\sim& 34 \left(\frac{\Delta t}{5\,\mathrm{yr}}\right)\left(\frac{C}{1}\right)\left(\frac{f_b}{0.01}\right)^{-1}\left(\frac{d}{\mathrm{Gpc}}\right)^{3},
\end{eqnarray}
where $\Delta t$ is the transient duration, $\Delta \Omega\approx0.82$ is the VLASS sky coverage fraction, $C$ is the completeness of our search, $f_b$ is the beaming factor, and $d$ is the distance. Out to a distance of 1\,Gpc (roughly z=0.2), we expect roughly a few dozen relativistic TDEs assuming a complete search and reasonable parameters $f_b\sim0.01$ and $\Delta t \sim 5\,\mathrm{yr}$. On the other hand, few transient candidates were identified at $z\lesssim0.2$. The smaller number of observed candidates is not unexpected, because our search is likely incomplete (i.e., $C<1$) due to additional criteria such as the selection of inverted spectra and exclusion of any host galaxies detectable in NVSS. This comparison implies that rarity does not disfavor the relativistic TDE interpretation for our transient candidates, but also that our sample of candidates does not provide useful constraints on the rate of radio-selected relativistic TDEs (especially when AGNs may exist in our sample). 

For the specific candidate VT\,J0243-2840, \citet{Somalwar2023J0243} obtained multi-frequency radio follow-up observations and found the radio SEDs to be consistent with an energetic outflow in an ordinary environment (electron density $\gtrsim$1\,cm$^{-3}$) that was mildly relativistic on average ($0.1c<\langle v \rangle < 0.6c$). VT\,J0243-2840 was interpreted as a candidate relativistic TDE by \citet{Somalwar2023J0243}, and thus would be an example supporting the relativistic TDE interpretation of our transient candidates. We note that there is ambiguity in this case (and likely in all our candidates) that the brightening could be due to some unusual AGN activity (see Section \ref{subsubsec:AGNVar} above). For example, \citet{Kunert-Bajraszewska2025} argue against the TDE interpretation for VT\,J0243-2840, VT\,J0535+8345, and VT\,J2239-2231, based on differences between these radio transients and known radio flares from thermal TDEs \citep[][]{Cendes2023,Somalwar2023TDEI}. However, based on the rate analysis above and the lack of strong AGN signatures and variability for our transient candidates with well-detected optical and IR counterparts, we still consider relativistic TDE as the most plausible transient interpretation. Follow-up observations will be needed to disentangle the two possibilities.

We do note, however, that a potential difference between our transient candidates and known relativistic TDEs is the timescale. Compared to relativistic TDEs (mostly the well-studied case of Swift\,J1644+57) that are seen to vary by a factor of a few over several hundred days, our transient candidates show little variability between the two epochs of VLASS ($\sim$500-900\,d in the rest frame). It is possible to reconcile this difference if epoch 1 and 2 of VLASS separately captured the rising and fading portions, respectively, but this scenario is unlikely to apply to all candidates. For example, the three candidates that have light curves from ASKAP show relatively smooth brightening over years (Figure~\ref{fig:ASKAPLC}). According to the simulation by \citet{Metzger2015}, only $\sim$50\% of TDEs similar to Swift\,J1644+57 detected in VLASS are expected vary by less than a factor of 2 between two epochs. Therefore, to explain such low levels of variability, our transient candidates likely evolve much slower than known relativistic TDEs selected in the X-ray and optical. 

\citet{Somalwar2023J0243} similarly remarked on the unusually long timescale of VT\,J0243-2840, which was brightening for $t/(1+z)\gtrsim1200\,\mathrm{d}$ (possibly $\gtrsim$3000\,d). VT\,J0243-2840 also had a break frequency ordering of $\nu_a < \nu_m < \nu_c$ in the most recent radio SED, where $\nu_a$ is the self-absorption frequency, $\nu_m$ is the characteristic frequency of electrons with minimum energy, and $\nu_c$ is the cooling frequency. Such ordering is expected during the earlier phase of the slow cooling scenario and should transition to $\nu_m < \nu_a$ at late times \citep[][]{Granot&Sari2002}, which happened for Swift\,J1644+57 at $t/(1+z)\gtrsim222\,\mathrm{d}$. \citet{Somalwar2023J0243} suggested that continual energy injection or low energy dissipation may explain the slow evolution.

In summary, we argue that relativistic TDEs are a plausible explanation for our transient candidates, especially if AGN activity can be ruled out in the host galaxies. However, compared to known on-axis relativistic TDEs, our candidates appear to evolve much more slowly over time. The difference in timescale is not necessarily a counterargument against the TDE interpretation, but rather reflects the complexity of TDEs, especially in the radio. Complex behaviors were seen in recent radio observations of thermal TDEs, which showed that a large fraction of optically-selected TDEs experience delayed radio brightening after $t\sim10^{2}-10^{3}\,\mathrm{d}$ \citep[][also see individual cases reported by \citealt{Horesh2021ASASSN15oi,Horesh2021iPTF16fnl,Cendes2022,Perlman2022,Goodwin2023,Christy2024,Zhang2024}]{Cendes2023}. Proposed explanations for the delayed brightening include delayed outflows \citep[][]{Horesh2021ASASSN15oi,Cendes2023} and off-axis jets \citep[][]{Matsumoto2023,Sfaradi2024}. For relativistic TDEs, some recent theoretical studies have suggested that precession can result in effectively weaker jets that become choked, explaining the rarity of relativistic TDEs, and break out of some chocked jets may lead to delayed radio brightening \citep[][]{Teboul2023,Lu2023}. At this point, many open questions exist regarding the exact late-time evolution and behavior of TDEs in the radio, and our candidates may occupy a mostly unexplored slow-evolving regime of relativistic TDEs. 

Finally, we note that TDEs studied so far have been almost exclusively selected in the X-ray and optical, but discoveries of radio-selected TDEs will bring forth a different perspective. For example, \citet{Somalwar2023TDEI} recently found tentative trends that radio-selected TDEs have fainter optical emission and prefer galaxies with lower stellar and black hole masses. \citet{Dykaar2024} began constraining volumetric rates of on-axis and off-axis relativistic TDEs through an untargeted search in the VAST pilot survey. Therefore, if (at least some of) our transient candidates are relativistic TDEs, the radio perspective can provide unique insights for the complex processes involved in these rare events. Follow-up observations and detailed analyses will be crucial for this purpose, which we leave for future works.

\subsubsection{Hypernebulae}\label{subsubsec:HNe}

Hypernebulae are nebulae inflated by hyperaccreting binary systems that contain a compact object and a donor star \citep[][]{Sridhar2022HNe,Sridhar2024HNev}. The extreme accretion is the result of unstable runaway mass transfer that occurs briefly before a stellar merger or a common-envelope event, which can exceed the Eddington accretion rate by orders of magnitude. Such super-Eddington accretion drives powerful winds and jets that produce luminous radio emission. While hypernebulae are mainly considered in the context of being potential progenitors to fast radio bursts \citep[e.g.,][]{Dong2024PRS,Ibik2024PRS}, they can also appear as slowly evolving radio transients that maintain inverted spectra at $\sim$GHz frequencies for years and millennia. Furthermore, based on the model established by \citet{Sridhar2022HNe}, the radio emission from hypernebulae can reach $L_{\nu}\gtrsim10^{30}\,\mathrm{erg}\,\mathrm{s}^{-1}\,\mathrm{Hz}^{-1}$ at 3.0\,GHz. Thus, hypernebulae are also a plausible explanation for our transient candidates.

However, we note that the observed positions of our transient candidates are concentrated at centers of galaxies. Such a preference is expected for nuclear transients (such as TDEs) but not for massive stellar binaries. Therefore, based on this preference, we argue that our sample of transient candidates is unlikely dominated by hypernebulae. Nonetheless, our sample could still contain some hypernebulae, which may be distinguishable through additional observations. For example, hypernebulae are expected to brighten and remain compact ($\lesssim$pc size) for many decades \citep[][]{Sridhar2022HNe}, and such properties can be tracked through continual monitoring and very long baseline interferometry (VLBI) observations. We leave this consideration for future works.

\subsection{Physical Constraints From Inverted Spectra}\label{subsec:physicsinvertedspectrum}

\begin{figure*}
\epsscale{1.14}
\plotone{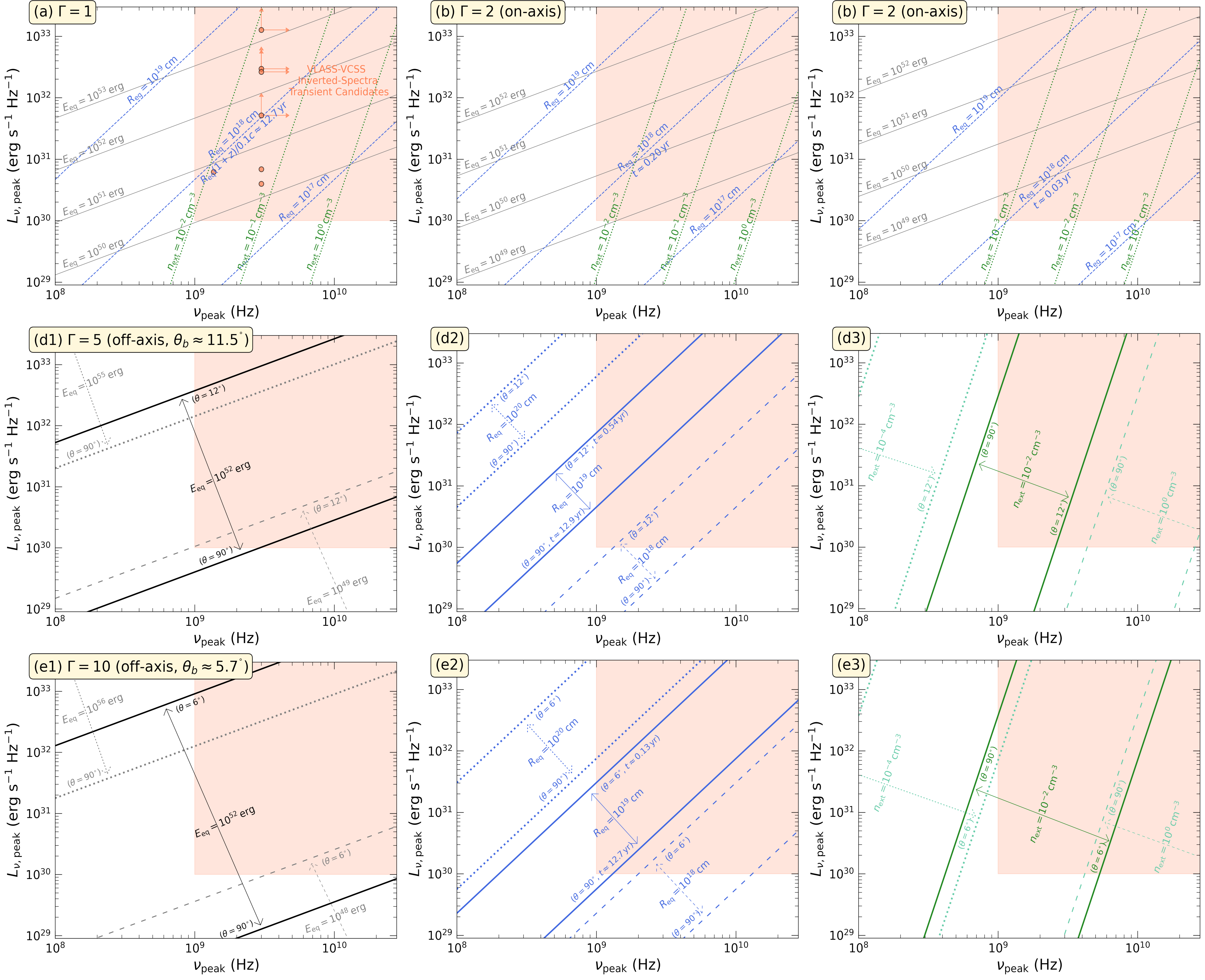}
\caption{Plots of $L_{\nu,\mathrm{peak}}$ vs. $\nu_{\mathrm{peak}}$ showing physical quantities derived following \citet{BarniolDuran2013} and \citet{Matsumoto2023} for various scenarios. The quantities, $E_{\mathrm{eq}}$, $R_{\mathrm{eq}}$, and $n_{\mathrm{ext}}$, are shown as lines of constant values in gray, blue, and green, respectively. The derived time $t$ for one value of $R_{\mathrm{eq}}$ is also shown for reference in each scenario (which scales linearly to other values of $R_{\mathrm{eq}}$). The approximate region of phase space covered by our transient candidates is shaded in orange (but note that it can extend towards the top right). In the top left panel, we also show crude locations of our transient candidates that have a photometric redshift and an inferred spectral shape. Peak flux density measurements (directly from surveys) are shown as circles for candidates with peaked spectra, while lower limits are indicated by arrows for candidates with highly inverted spectra. The label for each panel is placed at the top left with a description of the scenario. Panel (a) describes a non-relativistic outflow with $\Gamma = 1$. Panel (b) and (c) describe on-axis relativistic outflows with $\Gamma=2$ and $\Gamma=5$, respectively. Panels (d1), (d2), and (d3) describe the same off-axis outflow with $\Gamma=5$ (and $\theta_b = \cos^{-1}\beta$ is the boundary dividing on-axis and off-axis cases) but show the three physical quantities separately for clarity. For the off-axis case, a constant quantity spans a range that corresponds to the range of $\theta$. In our plots, we show three ranges spanned by arrows (labeled by the value of the quantity) and bounded by two lines with limiting values of $\theta$ (labeled at the left/right of the arrow for the lower/upper boundary). Only one boundary is shown with a lighter color when the other boundary is beyond the limits of our plots. Panels (e1), (e2), and (e3) are the same as panels (d1), (d2), and (d3), but with $\Gamma=10$.} \label{fig:PhysicalConstraints}
\end{figure*}

The theme of this study is transient selection based on an \textit{inverted spectrum}, motivated by the idea that this spectrum limits the brightness and frequency at the spectral peak and correspondingly constrains specific physical properties. We can now use the observed radio SEDs of our transient candidates to broadly examine the obtainable physical constraints, which will help us better understand what can be expected from a transient search that incorporates inverted spectrum as a selection criterion.

The radio SEDs of our transient candidates roughly constrain the peak frequency\footnote{For the spectral index to be $\alpha > 0$ between two points at 340\,MHz and 3.0\,GHz, assuming the simplest case of two intersecting power laws (no curvature) with $\alpha = 2.5/-0.75$ below/above the peak (standard for synchrotron spectrum), we expect $\nu_{\mathrm{peak}}>562\,\mathrm{MHz}$. Therefore, within an order of magnitude, our inverted spectra suggest $\nu_{\mathrm{peak}}\gtrsim\mathrm{GHz}$.} to $\nu_{\mathrm{peak}}\gtrsim\,\mathrm{GHz}$ and the peak luminosity to $L_{\nu,\mathrm{peak}}\gtrsim10^{30}\,\mathrm{erg}\,\mathrm{s}^{-1}\,\mathrm{Hz}^{-1}$ given that the VLASS measurements are near or below the spectral peak. We assume the emission to be synchrotron radiation and derive physical quantities that are a function of these two observables following the equipartition analysis of \citet{BarniolDuran2013} and \citet{Matsumoto2023}. Specifically, we calculate the equipartition energy $E_{\mathrm{eq}}$, radius $R_{\mathrm{eq}}$, and ambient electron density $n_{\mathrm{ext}}$ for both on-axis \citep[][]{BarniolDuran2013} and off-axis \citep[][]{Matsumoto2023} cases, adopting certain values of the Lorentz factor $\Gamma$ and viewing angles $\theta$ for illustration. We similarly characterize the outflow geometry using the area filling factor $f_A$ and the volume filling factor $f_V$. They are defined as $f_A \equiv A/(\pi R^2/\Gamma ^2)$ and $f_V \equiv V/(\pi R^3/\Gamma^4)$, where $R$, $A$, and $V$ are the radius, area, and volume, respectively.

However, because we have very limited observational constraints, we make a number of assumptions for our calculations. These assumptions are (i) equipartition between electrons and magnetic field, (ii) filling factors $f_V = 4/3$ and $f_A =1$ (spherical geometry) for the non-relativistic ($\Gamma =1$) case and $f_V = f_A = 1$ for relativistic ($\Gamma>1$) cases \citep[corresponding to a jet with a half-opening angle $\theta_j \approx 1/\Gamma$;][]{BarniolDuran2013}, (iii) density calculated by $n_{\mathrm{ext}}=N_e/(4\Gamma^2 V)$ where $N_e$ is the number of electrons and $V=f_V\pi R_{\mathrm{eq}}^3/\Gamma^4$ is the volume, (iv) $\eta=1$ but no correction to electron energy for $\nu_m < \nu_a$ (essentially ignoring this factor), (v) a redshift $z = 0.2$ (the approximate median redshift of our sample), and (vi) bulk of the total energy contained in hot protons leading to a correction factor of $\xi^{1/17}$ and $\xi^{11/17}$ to $R_{\mathrm{eq}}$ and $E_{\mathrm{eq}}$, respectively, where $\xi = 1+\epsilon_e^{-1} = 11$ with $\epsilon_e = 0.1$. In the non-relativistic case, we also apply a factor of $4^{1/17}$ and $4^{11/17}$ to $R_{\mathrm{eq}}$ and $E_{\mathrm{eq}}$, respectively, as discussed by \citet{BarniolDuran2013}. 

Before discussing the physical constraints, we emphasize that our assumptions are not always accurate or appropriate. Differing conditions such as deviation from equipartition, $\eta > 1$, $z\neq 0.2$, and different jet structures (e.g., narrow jets with $f_A = f_V = (\theta_j\Gamma)^2$ and $\theta_j < 1/\Gamma$) will change the exact values of the physical quantities. For example, if we assume $z = 1.0$ (largest photometric redshift of our sample) while keeping everything else the same, the derived $R_{\mathrm{eq}}$ and $E_{\mathrm{eq}}$ decrease by approximately a factor of 2 and 3, respectively, while $n_{\mathrm{ext}}$ increases by a factor of 3. Examination of all possible variations in the parameters is beyond the scope of this study, and the numbers we show should be treated (at best) as order-of-magnitude estimates meant to lead the discussion. 

In Figure~\ref{fig:PhysicalConstraints}, we show plots of $L_{\nu,\mathrm{peak}}$ vs. $\nu_{\mathrm{peak}}$ including lines of constant values of the derived physical quantities for various scenarios (labeled at the top left of each panel). The approximate region covered by our transient candidates is also shaded in orange for reference (but note that the shown borders are rough lower limits and the region can extend beyond the plot). In all scenarios, the dependence of the physical quantities is clear: a larger $L_{\nu,\mathrm{peak}}$ indicates a larger $E_{\mathrm{eq}}$, a larger $R_{\mathrm{eq}}$, and a smaller $n_{\mathrm{ext}}$, while a larger $\nu_{\mathrm{peak}}$ indicates a smaller $E_{\mathrm{eq}}$, a smaller $R_{\mathrm{eq}}$, and a larger $n_{\mathrm{ext}}$. Now, we examine the physical constraints associated with the inverted spectra of our transient candidates in each scenario and discuss potential implications.

\underline{\textit{Non-relativistic outflow}}: In panel (a) of Figure~\ref{fig:PhysicalConstraints}, we show the non-relativistic scenario. To produce the observed $L_{\nu,\mathrm{peak}}\sim10^{30}-10^{33}\,\mathrm{erg}\,\mathrm{s}^{-1}\,\mathrm{Hz}^{-1}$ at $\nu_{\mathrm{peak}}\sim1-10\,\mathrm{GHz}$ (orange region), a non-relativistic ($\Gamma=1$) outflow requires $E_{\mathrm{eq}}\sim10^{50}-10^{53}\,\mathrm{erg}$ at $R_{\mathrm{eq}} \sim 10^{17}-10^{19}\,\mathrm{cm}$ with $n_\mathrm{ext} \sim 1\,\mathrm{cm}^{-3}$. This energy range is quite high, which is a direct result of the high luminosity over the low GHz frequency range. Unsurprisingly, it is consistent with (or even larger than) energies inferred from relativistic transients such as GRB afterglows and relativistic TDEs \citep[][]{Beniamini2017,Eftekhari2018Sw1644} and is (up to orders of magnitude) beyond those inferred from SNe, LFBOTs, and thermal TDEs \citep[][]{Coppejans2020,Ho2022,Christy2024}. Such high energy simply reaffirms that we expect mainly the brightest and most energetic (extragalactic) transients to be selected in VLASS and VCSS based on inverted spectra because of the sensitivity limit (i.e., the sensitivity of VCSS requires a $\gtrsim15\,\mathrm{mJy}$ detection in VLASS). 

The emission size is related to the timescale and the outflow velocity of the transient. For relativistic transients, we can crudely estimate the timescale assuming that the outflow velocity at the non-relativistic phase can be represented by an average velocity that is some fraction of the speed of light \citep[which was the case for VT\,J0243-2840;][]{Somalwar2023J0243}. For the range $R_{\mathrm{eq}} \sim 10^{17}-10^{19}\,\mathrm{cm}$, we estimate a rough timescale of $t\approx R_{\mathrm{eq}}(1+z)/\langle v \rangle \lesssim 1-100\,\mathrm{yr}$ for $\langle v \rangle \gtrsim 0.1c$ (over the orange region). This range of emission size (and timescale) also overlaps with a density range of $n_{\mathrm{ext}}\sim10^{-2}-10^{0}\,\mathrm{cm}^{-3}$, and a smaller emission size generally corresponds to a larger density. 

At the fainter end, our transient candidates with $L_{\nu,\mathrm{peak}}\lesssim10^{30}-10^{31}\,\mathrm{erg}\,\mathrm{s}^{-1}\,\mathrm{Hz}^{-1}$ peaking at more than a few GHz can be explained by relatively old non-relativistic outflows with reasonable energies of $E_{\mathrm{eq}}\lesssim 10^{51}\,\mathrm{erg}$. These have been expanding for less than about a decade and are now in low density environments with $n_{\mathrm{ext}}\lesssim1\,\mathrm{cm}^{-3}$, which is also reasonable considering the large presumed distance from the galaxy centers \citep[for comparison with circumnuclear densities inferred from TDEs, see Figure 9 in][]{Christy2024}. This result suggests that at this luminosity, it is possible to use an inverted spectrum in VLASS and VCSS to restrict the timescale of a transient at the non-relativistic phase to less than a decade old. If the spectral peak can be constrained to higher than a few GHz, perhaps through in-band spectral indices (e.g., a large positive $\alpha$ within the VLASS band), it may be possible to select less energetic but even younger radio transients at an early stage of the non-relativistic phase. Therefore, in theory, an inverted spectrum from just one epoch of observation can be used to identify and motivate follow-up observations of radio transients that are potentially still young. 

We note that towards the brighter end ($L_{\nu,\mathrm{peak}}\gtrsim10^{33}\,\mathrm{erg}\,\mathrm{s}^{-1}\,\mathrm{Hz}^{-1}$), the inferred physical quantities can become unreasonable at low frequencies. Aside from the extremely high energy ($E_{\mathrm{eq}}\gtrsim 10^{53}\,\mathrm{erg}$, approaching or exceeding the energy budget of a normal TDE), the timescale in this regime can be longer than many decades, which would be inconsistent with the NVSS non-detection two decades ago. This could potentially be a problem for our three candidates with the highest estimated spectral luminosities -- VT\,J0013-3125, VT\,J1400+6413, VT\,J1757+2929 with $L_{3.0\mathrm{GHz}}\gtrsim10^{32}-10^{33}\,\mathrm{erg}\,\mathrm{s}^{-1}\,\mathrm{Hz}^{-1}$. However, they have $\nu_{\mathrm{peak}}\gtrsim 3.0\,\mathrm{GHz}$ (see Figure~\ref{fig:RadioSEDs} and Figure~\ref{fig:PhysicalConstraints}), where the inferred energy and emission size can be much smaller and more reasonable. If there is inconsistency, it could be an indication that our model assumptions are not appropriate (e.g., the outflow may be relativistic with a smaller energy and timescale; see discussions of other panels of Figure~\ref{fig:PhysicalConstraints} below) or an AGN origin is a more preferable explanation than a transient event at such high luminosity (perhaps supporting the case of highly inverted variable PS source discussed in Section \ref{subsubsec:AGNVar}). We are unable to make any conclusive arguments for either case considering
model degeneracy and our limited observational constraints.

Lastly, we note that in theory, an inverted spectrum can provide strong constraints on the ambient density of radio transients. Unfortunately, the setup of VLASS and VCSS does not have sufficient depth or high enough frequency that would be useful for selecting radio transients in high density environments \citep[such as the discovery by][]{Dong2021}. Nonetheless, the prospect of such utility of an inverted spectrum should be considered in future surveys that have much greater capabilities.

\underline{\textit{On-axis relativistic outflow}}: In panel (b) and (c) of Figure~\ref{fig:PhysicalConstraints}, we show the on-axis relativistic scenario. For this scenario, we consider two values of Lorentz factor, $\Gamma = 2$ and $\Gamma = 5$, motivated by the early Lorentz factor inferred from Swift\,J1644+57 \citep[][]{Eftekhari2018Sw1644}. We calculate the timescale using $t \approx R_{\mathrm{eq}}(1-\beta)(1+z)/(\beta c)$, where $\beta = v/c$ \citep[][]{BarniolDuran2013}. Compared to the non-relativistic case, lower $E_{\mathrm{eq}}$ is needed to produce the same brightness and the timescale is now much smaller at the same $R_{\mathrm{eq}}$, especially with larger values of $\Gamma$. At the same peak luminosity and frequency, a larger $\Gamma$ also leads to a larger $R_{\mathrm{eq}}$ and a smaller $n_{\mathrm{ext}}$.

At high luminosities, the inferred $E_{\mathrm{eq}}$ and timescale is much more reasonable compared to the non-relativistic case. In general, it appears that our transient candidates can reasonably be explained by on-axis relativistic outflows. In this case, an inverted spectrum in theory provides an even stronger constraint on the age of the transient. However, we note that if the outflow was relativistic during VLASS epoch 1, it should have been decelerating over time in order to explain the observed inverted spectrum in epoch 2. If the outflow was not decelerating (constant $\Gamma$), at a constant $E_{\mathrm{eq}}$, the expansion over $\gtrsim$3\,yr (leading to more than an order of magnitude increase in $R_{\mathrm{eq}}$ for $\Gamma=2$) should result in a significant decrease in $\nu_{\mathrm{peak}}$, which was not observed in epoch 2. Alternatively, for a constant $\Gamma$, it may be possible to maintain an inverted spectrum if $E_{\mathrm{eq}}$ increases by multiple orders of magnitude (perhaps through energy injection). However, the increase in energy would likely result in noticeable brightening, which was also not observed in epoch 2 for any of our transient candidates. Therefore, the natural explanation for the inverted spectrum observed in both VLASS epochs is deceleration of the outflow.

In the on-axis relativistic scenario, we note that the inferred ambient density can be quite low, especially for large values of $\Gamma$. For $\Gamma=5$, at a high luminosity and low frequency (within the orange region), the density becomes $n_{\mathrm{ext}}\lesssim 10^{-3}\,\mathrm{cm}^{-3}$, which perhaps is explainable by the hot ionized medium or dwarf galaxies devoid of gas \citep[e.g.,][]{Putman2021}. This density would be even lower for larger $\Gamma$, which could be an indication that such on-axis relativistic outflow cannot reasonably explain our transient candidates.

\underline{\textit{Off-axis relativistic outflow}}: In panels (d) and (e) of Figure~\ref{fig:PhysicalConstraints}, we show the off-axis scenario. We consider two values of Lorentz factor, $\Gamma=5$ and $\Gamma=10$, with the emission beamed away from the line of sight, i.e., viewing angles $\theta > \theta_b$ (up to 90$^{\circ}$), where $\theta_b = \cos^{-1}\beta$ is the boundary between on-axis and off-axis angles \citep[following][]{Matsumoto2023}. We calculate the timescale using $t \approx R_{\mathrm{eq}}(1-\beta\cos\theta)(1+z)/(\beta c)$, leading to larger $t$ for larger $\theta$. As expected, a higher energy is required to produce the same luminosity at a larger viewing angle. In general, at small viewing angles, the difference between the on-axis and off-axis case is small, and the inferred values of $E_{\mathrm{eq}}$, $R_{\mathrm{eq}}$, and $n_{\mathrm{ext}}$ are still reasonable, at least for small $\Gamma$.

At small values of $\Gamma$ and relatively low luminosities, our transient candidates can also be reasonably explained by the edge-on ($\theta=90^{\circ}$) case. For example, for $\Gamma=5$ and $\theta=90^{\circ}$, transients with $L_{\nu,\mathrm{peak}}\sim10^{30}-10^{31}\,\mathrm{erg}\,\mathrm{s}^{-1}\,\mathrm{Hz}^{-1}$ and $\nu_{\mathrm{peak}}\sim1-10\,\mathrm{GHz}$ can be explained by an outflow with $E_{\mathrm{eq}}\sim10^{52}\,\mathrm{erg}$, $R_{\mathrm{eq}}\sim10^{18}\,\mathrm{cm}$ ($t\sim1\,\mathrm{yr}$), and $n_{\mathrm{ext}}\sim 10^{-1}-10^{1}\,\mathrm{cm}^{-3}$. However, at higher luminosities of $L_{\nu,\mathrm{peak}}\gtrsim10^{32}\,\mathrm{erg}\,\mathrm{s}^{-1}\,\mathrm{Hz}^{-1}$, the edge-on case requires an extremely high $E_{\mathrm{eq}}\sim10^{54}\,\mathrm{erg}$, which is potentially unphysical for an ordinary TDE. This problem is exacerbated at larger values of $\Gamma$, where larger $E_{\mathrm{eq}}$ is needed to produce the same luminosity. Thus, our transient candidates are less likely explained by off-axis outflows at large viewing angles moving at extreme velocities.

We note that the issue of low $n_{\mathrm{ext}}$ at high $\Gamma$ also exists in the off-axis case, particularly for small viewing angles. For example, for $\Gamma=10$ and a peak frequency in the low GHz range, the inferred ambient density is $n_{\mathrm{ext}} \lesssim 10^{-4}-10^{-3}\,\mathrm{cm}^{-3}$ at $\theta\approx \theta_b$. Thus, this again could be an indication that relativistic outflows with large $\Gamma$ close to on-axis cannot reasonably explain our transient candidates.

In general, for the off-axis case, an inverted spectrum similarly constrains the possible $E_{\mathrm{eq}}$, $R_{\mathrm{eq}}$, and $n_{\mathrm{ext}}$. However, with the introduction of $\theta$ and the resulting degeneracy, the constraints in this case is weaker (i.e., at a single point of ($\nu_{\mathrm{peak}}$, $L_{\nu,\mathrm{peak}}$), possible values of the physical parameters can span orders of magnitude over the range of $\theta$). Unfortunately, breaking this degeneracy is not easy and would require additional inputs such as apparent velocity measured from VLBI observations \citep[][]{Matsumoto2023}. If the viewing angle is unknown, the effectiveness of using inverted spectra for constraining physical parameters is limited.

\subsection{Feasibility of Search Method}\label{subsec:feasibilitymethod}

In this final discussion, we briefly reflect on the feasibility of our method for searching and selecting inverted-spectra transients, and consider caveats and possible improvements for future searches. Our method of transient search is based mainly on the non-detection in NVSS and detection in VLASS above a certain threshold. The two surveys span a temporal baseline of more than two decades, and since radio transients generally have much shorter timescales than this \citep[][]{Pietka2015}, our search should in theory capture the instantaneous number of transients on the sky above the flux density threshold and associated with a (previously) radio-quiet host. After introducing additional criteria on compactness and spectral index and removing cataloged AGNs, we ended up with a sample of 21 inverted-spectra transient candidates. Conveniently, because of our high flux density threshold (15\,mJy limited by the sensitivity of VCSS), VLASS detections are on average $\gtrsim100\sigma$ and we are significantly less troubled by false positives due to random noise structures. However, the down side of the high threshold and additional search criteria is that our final sample of transient candidates is fairly small compared to the thousands of variable and transient sources that have been found in VLASS (D. Dong, private communication).

From RACS-mid measurements, we found that nine of our candidates have brightened by $\gtrsim$300\%--4000\% at 1.4\,GHz after NVSS, confirming their transient nature (including extreme AGN brightening) and demonstrating the successful side of our transient search. However, seven of our candidates have RACS-mid measurements that are only $\sim$20\%--80\% brighter than the NVSS upper limits. For these candidates, (even if the chance may be slim) we cannot rule out extremely rare cases of highly inverted variable PS sources that were slightly fainter than the NVSS upper limit but happened to brighten moderately in VLASS and RACS-mid. Although these PS sources are interesting objects in themselves, they are not considered transient events. ``Contamination'' by such sources is possible because we have compared surveys at different frequencies, i.e., VLASS (3.0\,GHz) and NVSS (1.4\,GHz), and a difference in flux density can still be explained by an extraordinarily large spectral index. We note that it is difficult to estimate the expected number of PS sources in our sample due to the lack of robust statistics on variable PS sources with highly inverted spectra. Long-term monitoring will help distinguish real transient events from variable PS sources, as transients should evolve substantially over time.

Altogether, we conclude that our search is in general a success in identifying inverted-spectra transients, with the caveat that some fraction of the identified candidates may be contaminated by extremely rare PS sources. In future searches, the impact of such contamination can be reduced by raising the flux density threshold for VLASS such that connecting VLASS and NVSS would result in an unphysical spectral index. It is also possible to completely avoid such contamination by comparing between epochs of VLASS at the same frequency of 3.0\,GHz (searches between epochs of VLASS is currently ongoing; D. Dong et al., in prep.). 

Also, we originally designed our method to be robust rather than complete, resulting in a fairly small sample of transient candidates. However, in retrospect, it is possible to adjust the search criteria for a more complete search and a larger sample size. For example, the compactness criterion may be relaxed and slightly extended sources can also be examined for the possibility of being transient. In addition, rather than defining transient behavior based on non-detections in NVSS, a variability criterion would be better, because a previous non-detection is fundamentally the same as having a lower limit on the variability. In other words, we can define a transient to be any variable source above a certain variability threshold, which crudely mimics how transients are found through difference imaging (a better method but hard to carry out in our case). We leave these considerations for future searches, especially for later epochs of VLASS and VCSS.

\section{Summary \& Conclusion} \label{sec:conclusion}

In this study, we presented our search for transients with inverted spectra ($\alpha > 0$) in epoch 1 of VLASS and VCSS. Our search resulted in 21 inverted-spectra transient candidates that are not associated with cataloged AGNs. To the best of our knowledge, none of our candidates have been reported as transients at other wavelengths. In searches of radio surveys, three (VT\,J0243-2840, VT\,J0535+8345, and VT\,J2239-2231) have previously been reported as radio transients in the literature \citep[][]{Somalwar2023J0243,Kunert-Bajraszewska2025}. Many of our candidates have optical and IR counterparts, of which nine are matched to the centers of galaxies (Figure~\ref{fig:OpticalImages}) that have photometric redshifts, while four candidates have no optical or IR counterparts. However, none of our candidates are matched to known stellar counterparts. We also searched for X-ray counterparts, which unfortunately yielded few results from available data and very limited constraints. 

For our transient candidates, we compiled radio measurements from recent and ongoing radio surveys and constructed crude radio SEDs (Figure~\ref{fig:RadioSEDs}). All of our candidates with available measurements also show inverted spectra in epoch 2 of VLASS and VCSS. With RACS measurements between VLASS and VCSS, assuming slow evolution over time, we found broadly two types of spectra over $\nu\sim0.8-3.0\,\mathrm{GHz}$: (i) highly inverted spectra with $\alpha \sim 2.0-3.0$ that suggest $\nu_{\mathrm{peak}}\gtrsim 3.0\,\mathrm{GHz}$ and (ii) peaked spectra that suggest $\nu_{\mathrm{peak}}\lesssim 1.0-3.0\,\mathrm{GHz}$. The RACS measurements also confirmed in almost all cases that our candidates experienced some level ($\gtrsim$20\%--4000\%) of brightening after NVSS at 1.4\,GHz. From the VLASS epoch 1 and 2 measurements, we found that most of our candidates experienced only marginal variability at 3.0\,GHz between the two epochs, with the majority not being statistically significant (Figure~\ref{fig:Variability}). However, from searching ASKAP surveys, we found light curves for three of our candidates that showed smooth brightening over hundreds of days and no evidence for any sporadic variation (Figure~\ref{fig:ASKAPLC}).

Based on the results of our search, we considered possible classifications for our inverted-spectra transient candidates. From the high spectral luminosity of $L_{3.0\mathrm{GHz}}\sim10^{30}-10^{33}\,\mathrm{erg}\,\mathrm{s}^{-1}\,\mathrm{Hz}^{-1}$ and marginal variability over $t/(1+z_{\mathrm{phot}})\gtrsim500-900\,\mathrm{d}$, we disfavored non-relativistic transients (including SNe, LFBOTs, and thermal TDEs) and GRB afterglows as plausible classifications for our candidates. We also disfavored in general Galactic transients due to the lack of any bright optical stellar counterparts, although we cannot completely rule out bright Galactic radio transients (e.g., classical novae) at large distances. We found the most plausible transient classification to be relativistic TDEs, especially considering the spatial coincidences with centers of galaxies. However, without optical spectral classifications and sufficient X-ray constraints, we were unable to rule out the presence of AGNs (presumably with weaker optical and IR emission) in any of the host galaxies. Thus, our transient candidates that brightened significantly (by $\gtrsim$300\%--4000\%) can still be explained by young jets launched by AGNs, while candidates that brightened at least marginally (by $\gtrsim$20\%--80\%) may be explainable by extremely rare cases of highly inverted variable PS sources. 

Finally, we examined physical constraints associated with the inverted spectra from VLASS and VCSS following the equipartition analyses of \citet{BarniolDuran2013} and \citet{Matsumoto2023}. We found that our transient candidates can reasonably be explained by non-relativistic outflows, on-axis relativistic outflows, or off-axis relativistic outflows, but with some differences in the inferred energies, emission sizes, timescales, and ambient densities. In general, a high energy of $E_{\mathrm{eq}}\gtrsim10^{49}-10^{53}\,\mathrm{erg}$ is required to produce the observed high luminosity. A peak frequency at $\nu_{\mathrm{peak}}\gtrsim1-10\,\mathrm{GHz}$ would constrain the ambient density to $n_{\mathrm{ext}}\gtrsim 10^{-3}-10^{0}\,\mathrm{cm}^{-3}$ and the outflow duration to months or years. We also noted that certain solutions could be unphysical for our transient candidates. For example, a high velocity (say $\Gamma > 10$) on-axis outflow could result in an unrealistically low ambient density ($n_{\mathrm{ext}}\lesssim 10^{-4}-10^{-3}\,\mathrm{cm}^{-3}$), and a high velocity outflow viewed edge on ($\theta=90^{\circ}$) could result in a prohibitively large energy ($E_{\mathrm{eq}}\gtrsim10^{54}-10^{56}\,\mathrm{erg}$). These solutions are perhaps an indication that our transient candidates are unlikely explainable by extremely fast outflows, particularly those viewed at large off-axis angles. 

Overall, with the current setup of VLASS and VCSS, inverted spectra can be a useful criterion for selecting luminous relativistic radio transients, with potential applications for transient searches in epoch 2 and 3 of VLASS. This would also provide some constraints on timescale, ranging from months to years, although this is highly dependent on factors that are typically unknown during a transient search, such as the outflow velocity. Unfortunately, at high luminosities, the constraint on peak frequency ($\nu_{peak}\gtrsim3\,\mathrm{GHz}$) is insufficient for distinguishing transients in high density environments. We also remarked that due to the degeneracy introduced by the viewing angle, the effectiveness of inverted spectra in constraining physical properties is limited. While there certainly is potential in the utility of spectral information in a transient search, its merit may perhaps be better realized in future surveys with much higher sensitivities conducted by next-generation facilities such as the SKA, the Deep Synoptic Array-2000, or possibly the next-generation VLA. 

\begin{acknowledgments}

We thank the anonymous reviewer for helpful comments that improved this manuscript. We thank Hannah Dykaar for catalog suggestions, Biny Sebastian and Yjan Gordon for discussions about the VLASS catalog, and Ron Ekers and Elaine Sadler for sharing AT20G snapshot images. Y.C. would also like to thank Alex Andersson, Navin Sridhar, and Ayush Pandhi for insightful comments about the manuscript. Y.C. acknowledges support from the Natural Sciences and Engineering Research Council of Canada (NSERC) Canada Graduate Scholarships – Doctoral Program. The Dunlap Institute is funded through an endowment established by the David Dunlap family and the University of Toronto. B.M.G. acknowledges the support of the Natural Sciences and Engineering Research Council of Canada (NSERC) through grant RGPIN-2022-03163, and of the Canada Research Chairs program. K.R. thanks the LSST-DA Data Science Fellowship Program, which is funded by LSST-DA, the Brinson Foundation, and the Moore Foundation; Their participation in the program has benefited this work.

Basic research in radio astronomy at the U.S. Naval Research Laboratory (NRL) is supported by 6.1 Base funding. Construction and installation of VLITE was supported by the NRL Sustainment Restoration and Maintenance fund. The VLA is operated by the National Radio Astronomy Observatory (NRAO). The NRAO is a facility of the National Science Foundation operated under cooperative agreement by Associated Universities, Inc. CIRADA is funded by a grant from the Canada Foundation for Innovation 2017 Innovation Fund (Project 35999), as well as by the Provinces of Ontario, British Columbia, Alberta, Manitoba and Quebec. This research has made use of the VizieR catalogue access tool, CDS, Strasbourg, France \citep{10.26093/cds/vizier}. The original description of the VizieR service was published in \citet{vizier2000}.

This scientific work uses data obtained from Inyarrimanha Ilgari Bundara / the Murchison Radio-astronomy Observatory. We acknowledge the Wajarri Yamaji People as the Traditional Owners and native title holders of the Observatory site. CSIRO’s ASKAP radio telescope is part of the Australia Telescope National Facility (https://ror.org/05qajvd42). Operation of ASKAP is funded by the Australian Government with support from the National Collaborative Research Infrastructure Strategy. ASKAP uses the resources of the Pawsey Supercomputing Research Centre. Establishment of ASKAP, Inyarrimanha Ilgari Bundara, the CSIRO Murchison Radio-astronomy Observatory and the Pawsey Supercomputing Research Centre are initiatives of the Australian Government, with support from the Government of Western Australia and the Science and Industry Endowment Fund. This paper includes archived data obtained through the CSIRO ASKAP Science Data Archive, CASDA (https://data.csiro.au).

\end{acknowledgments}

\vspace{5mm}
\facilities{VLA, ASKAP, LOFAR}

\software{Astropy \citep{2013A&A...558A..33A,2018AJ....156..123A}, TOPCAT \citep[][]{2005ASPC..347...29T}, STILTS \citep[][]{2006ASPC..351..666T}, NumPy \citep[][]{Harris2020Numpy}, Matplotlib \citep[][]{Hunter2007Matplotlib}, SciPy \citep{2020scipy}.}

\clearpage

\appendix

\section{VLASS Systematic Flux Density Underestimation}\label{apdx:FluxCal}

\begin{figure}
\epsscale{1.15}
\plotone{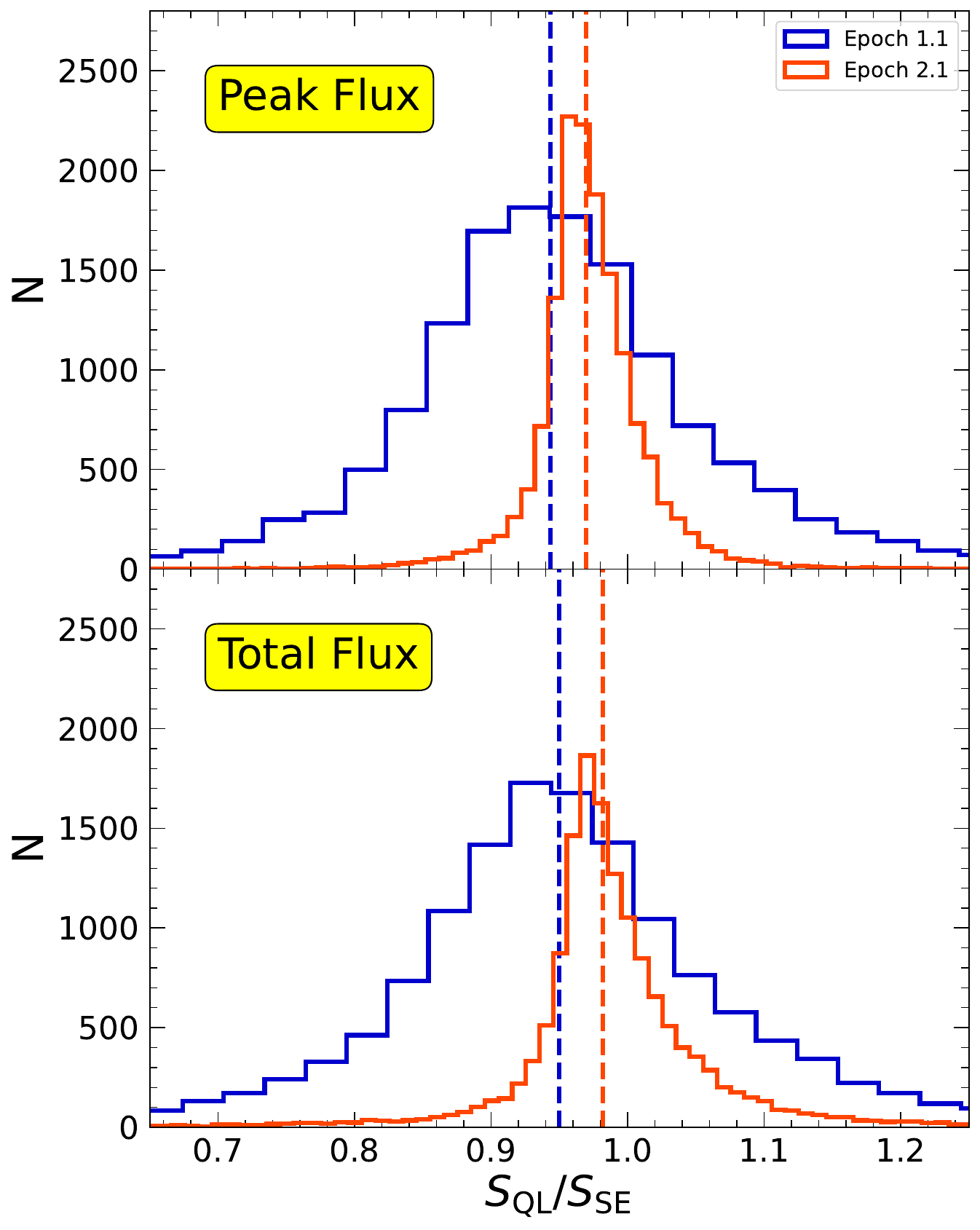}
\caption{The distributions of the ratios of flux densities between the VLASS QL and SE sources, $\mathrm{S}_{\mathrm{QL}}/\mathrm{S}_{\mathrm{SE}}$. QL flux densities from epoch 1.1 (blue) and 2.1 (orange) are shown separately. The top and bottom panels show the ratios of peak flux density and total flux density, respectively. The medians of the distributions are shown with dashed lines. The underestimation of flux density in QL epoch 1.1 is worse compared to QL epoch 2.1 with significantly larger spread in distribution.
\label{fig:FluxRatioQLSE}}
\end{figure}

\begin{deluxetable}{ccccc}
\tablecaption{Statistics of $\mathrm{S}_{\mathrm{QL}}/\mathrm{S}_{\mathrm{SE}}$\label{tab:FluxQLSEStats}}
\tablewidth{0pt}
\tablehead{
\colhead{Epoch} & \multicolumn{2}{c}{Peak Flux} & \multicolumn{2}{c}{Total Flux}
\\
\colhead{} & \colhead{Median} & \colhead{MAD} &  \colhead{Median} & \colhead{MAD}
}
\startdata
1.1 & 0.943 & 0.063 & 0.950 & 0.070     \\
2.1 & 0.969 & 0.018 & 0.982 & 0.026      \\    
\enddata
\tablecomments{MAD : median absolute deviation.}
\end{deluxetable}

Flux density measurements from the VLASS QL images are known to be systematically underestimated as a result of the imaging pipeline and residual phase and amplitude errors \citep[][]{Memo13,Gordon2021VLASS}. Additionally, this issue was worse for epoch 1.1 due to erroneous pointing offsets \citep[][]{Memo13}. Therefore, it is important to assess the systematic flux density underestimations in order to accurately compare the VLASS QL catalogs with other radio catalogs as well as between epochs. 

To quantify the systematic underestimations, we directly compared the CIRADA VLASS QL catalogs used in this study (version 3 and 2 for epoch 1 and 2, respectively) with the CIRADA VLASS Single Epoch (SE) catalog (version 2) that covers $\sim$1000\,deg$^{2}$ of the sky observed in epoch 2.1. The SE catalog was derived from higher quality VLASS images processed through the SE imaging pipeline with improved beam sampling, deeper cleaning, and self-calibration \citep[][]{Memo17}{}{}. The flux calibration accuracy of the SE images is known to be $\sim$3\%, and therefore is suitable as a reference.

For the comparison, to minimize the impact of contamination, we followed the recommendations of the catalog User Guide (February 1, 2023 version) to select cleaner samples of the QL and SE catalogs. Specifically, the cleaner sample contains components with $Quality\_flag=0|4|8|12$, $P\_sidelobe<0.05$, $S\_Code\neq\mathrm{E}$, and $Duplicate\_flag<2$. Furthermore, we restricted the samples to sources relevant to this study, i.e., compact sources with $S\_Code =\mathrm{S}$ and $Maj < 5.0''$ as well as flux densities above the completeness limit of 3\,mJy. 

With the final samples, we cross-matched the QL and SE sources using a match radius of $2.0''$ and calculated the flux density ratio $\mathrm{S}_{\mathrm{QL}}/\mathrm{S}_{\mathrm{SE}}$. Since the SE catalog covers a part of epoch 2.1, the matched QL sources are almost exclusively from epoch 1.1 and 2.1, allowing us to characterize the underestimations in these two epochs. Figure \ref{fig:FluxRatioQLSE} shows the the distributions of $\mathrm{S}_{\mathrm{QL}}/\mathrm{S}_{\mathrm{SE}}$ for both peak and total flux density. As expected, the underestimation of flux density in QL epoch 1.1 is worse than QL epoch 2.1 with a larger spread spread in distribution indicating significant uncertainty.

From these distributions, we derived the median and Median Absolute Deviation (MAD) values of $\mathrm{S}_{\mathrm{QL}}/\mathrm{S}_{\mathrm{SE}}$, given in Table \ref{tab:FluxQLSEStats}. The median values imply that the flux density is underestimated by $\sim$5\% in QL epoch 1.1 and $\sim$2\%--3\% in QL epoch 2.1. The MAD values indicate that the level of uncertainty is at $\sim$7\% in QL epoch 1.1 and $\lesssim$3\% in QL epoch 2.1. Note that although we could not make direct comparisons for QL epoch 1.2 and 2.2, the flux density accuracy at these epochs should be similar to QL epoch 2.1 \citep[][]{Memo13}. Therefore, for the VLASS QL catalogs used in this study, we chose to correct the total flux density measurements by a factor of $1/0.95$ for epoch 1.1 and by a factor of $1/0.98$ for epoch 1.2+2.1+2.2. In addition, we also added 7\% and 3\% of the total flux density measurements in quadrature to the uncertainties for epoch 1.1 and epoch 1.2+2.1+2.2, respectively. 

\section{Other Radio Images of Transient Candidates}\label{apdx:OtherRadioImg}

In Figure~\ref{fig:OtherRadioImg1} and Figure~\ref{fig:OtherRadioImg2}, we show cutout images for our transient candidates from WENSS, TGSS, VLASS epoch 2, VCSS epoch 2, RACS-low, RACS-mid, and LoTSS, with box sizes (side lengths) of 10\,arcmin, 5\,arcmin, 0.5\,arcmin, 5\,arcmin, 5\,arcmin, 3\,arcmin, 2\,arcmin, respectively. A blank image of zero values is shown if the candidate is not observed by a survey. All detections across different surveys appear to be consistent with a compact source. Note that we do not show FIRST images because all of our transient candidates are outside the FIRST coverage. We also checked existing snapshot images for the AT20G survey (R. Ekers \& E. Sadler, private communication) but found that our transient candidates are not within their coverage.

\begin{figure*}
\epsscale{1.0}
\plotone{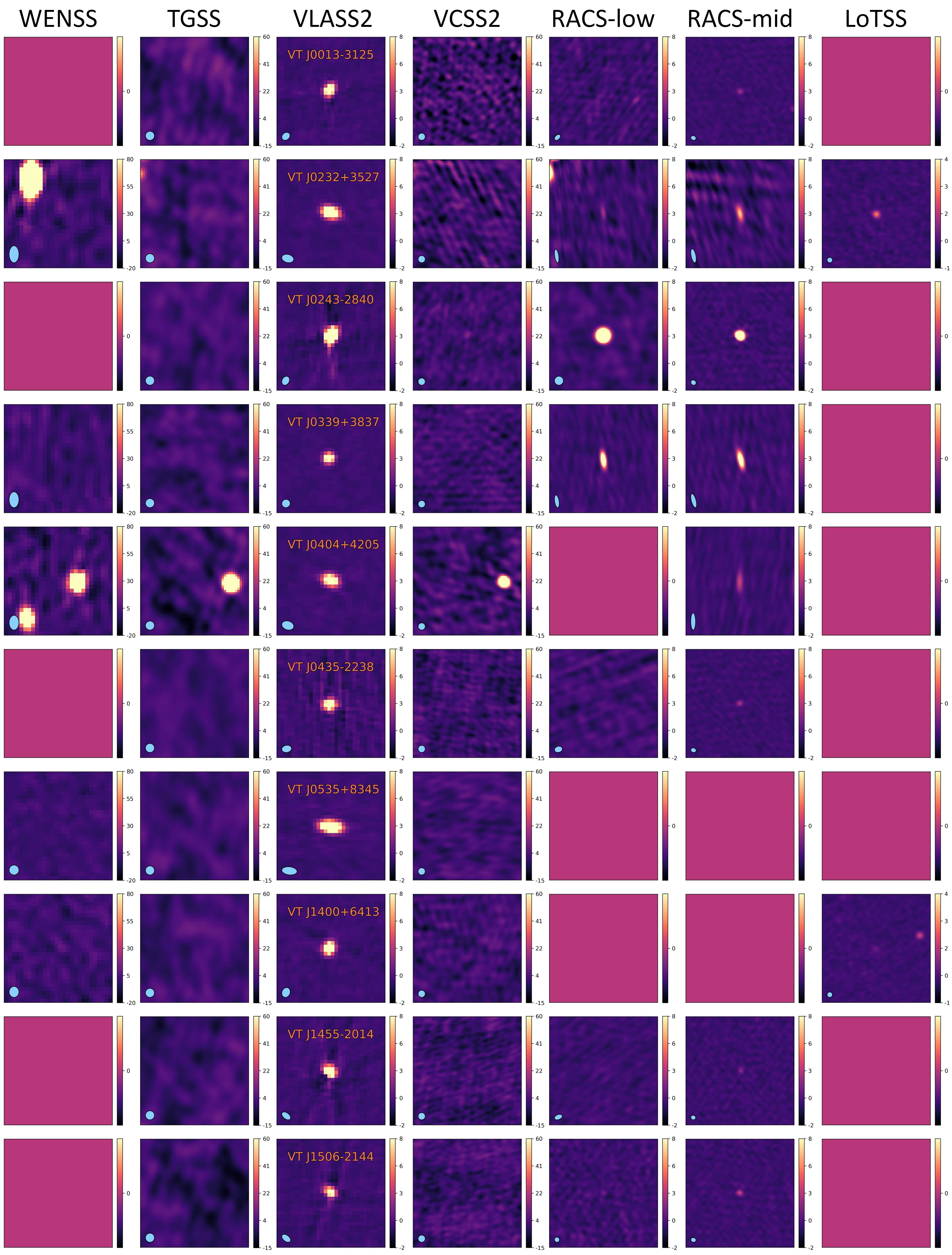}
\caption{Other radio images for our transient candidates (separated by rows). Names of surveys are shown at the top and names of the transient candidates are shown on the VLASS epoch 2 images. \label{fig:OtherRadioImg1}}
\end{figure*}

\begin{figure*}
\epsscale{1.0}
\plotone{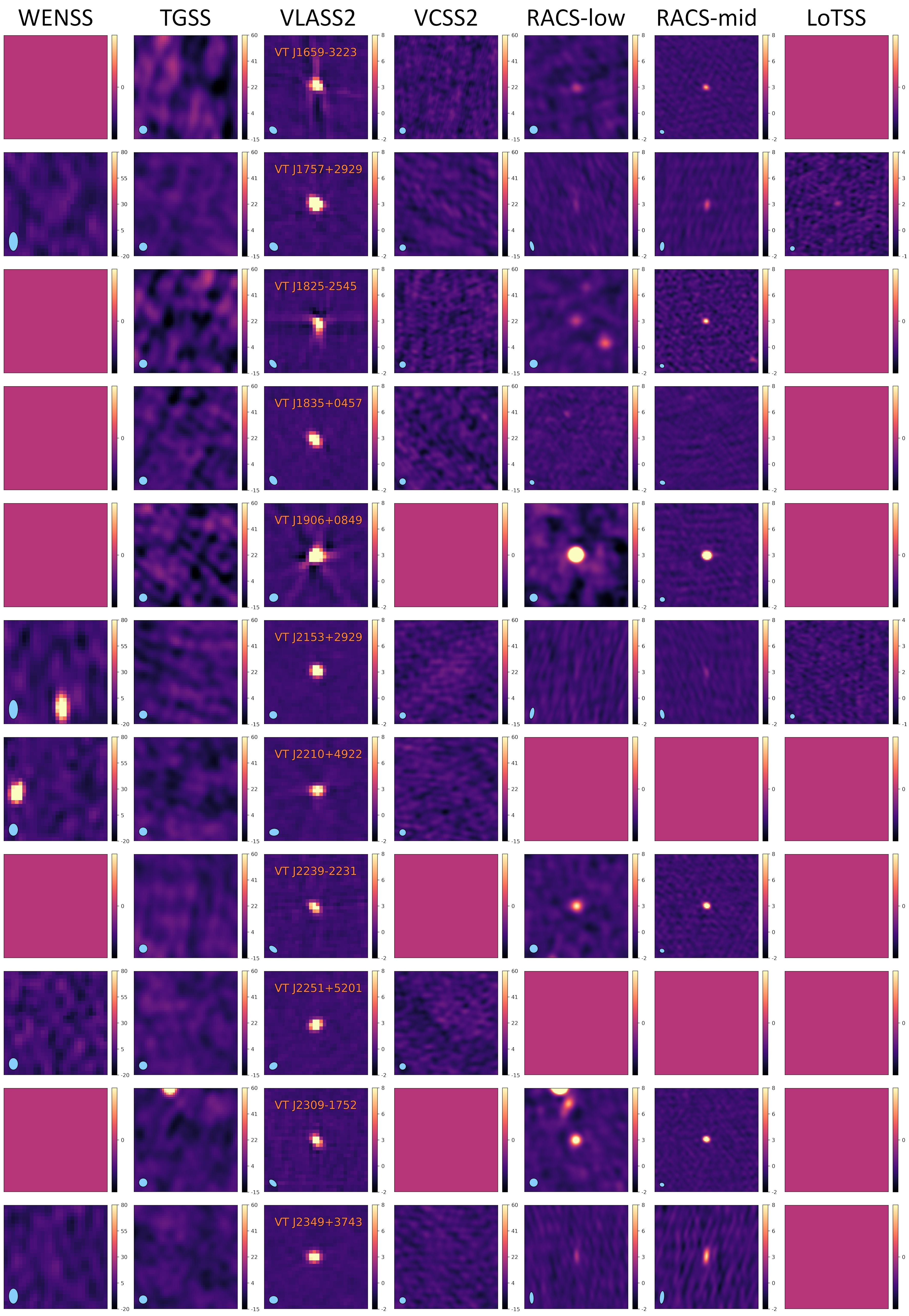}
\caption{Continuation of Figure~\ref{fig:OtherRadioImg1}. \label{fig:OtherRadioImg2}}
\end{figure*}

\bibliography{ref}{}
\bibliographystyle{aasjournal}

\end{document}